\documentclass[twocolumn]{aastex701}
\usepackage{amsmath}
\usepackage{xcolor}
\usepackage{booktabs}
\usepackage{gensymb}

\begin{document}

\title{The Dynamics of Planetary Ejection}

\author[orcid=0000-0002-7670-670X,sname='Rice']{Malena Rice}
\affiliation{Department of Astronomy, Yale University, 219 Prospect Street, New Haven, CT 06511, USA}
\email{malena.rice@yale.edu}

\author[0000-0003-1827-9399]{William DeRocco}
\affiliation{Maryland Center for Fundamental Physics, University of Maryland, College Park, 4296 Stadium Drive, College Park, MD 20742, USA}
\affiliation{Department of Physics \& Astronomy, The Johns Hopkins University, 3400 N. Charles Street, Baltimore, MD 21218, USA}
\email{derocco@umd.edu}

\author[0000-0001-8974-0758]{Sean N. Raymond}
\affiliation{Laboratoire d’astrophysique de Bordeaux, Univ. Bordeaux, CNRS, B18N, all\'{e}e Geoffroy Saint-Hilaire, F-33615 Pessac, France}
\email{rayray.sean@gmail.com}


\correspondingauthor{Malena Rice}
\email{malena.rice@yale.edu}

\begin{abstract}

The ubiquity of free-floating planets inferred from microlensing and direct imaging surveys suggests that planetary ejection---a process in which planets initially born encircling a stellar host become gravitationally unbound---is common. Four overarching mechanisms have been proposed to induce planetary ejection: close approaches of neighboring planets, instabilities in binary or multi-star systems, stellar and planetary flybys, and post-main-sequence stellar evolution. Here we review the mechanisms underlying planetary ejection, as well as predictions derived from each. Current and upcoming microlensing surveys offer the potential to test existing models and distinguish between planetary ejection mechanisms, providing further insight into the demographic-level architectures of exoplanets across stellar environments.
\end{abstract}




\section{Introduction}

The signatures of multi-scale dynamic environments are imprinted on today's census of observed planetary systems. Both theoretical and observational constraints have indicated that planets' evolved configurations are often not representative of the initial outcomes of planet formation \citep{papaloizou2006planet,davies2014long,morbidelli2018dynamical,raymond2020solar}. Instead, a given main-sequence (or post-main-sequence) system reflects the integrated sculpting of 
millions to billions of years of processing. Orbits may evolve substantially over a system's lifetime, undergoing shifts and instabilities. In many scenarios, these orbital shifts may result in planets being ejected from their host systems.

Numerical simulations indicate that planetary ejection is a common outcome of dynamical interactions between planets and their perturbers \citep{weidenschilling1996gravitational,rasio1996dynamical,lin1997on,chatterjee2008dynamical,ford2014architectures}. Observational constraints, too, point toward a high rate of planetary ejection: microlensing surveys have unveiled a vast population of planets orbiting within the Milky Way, unbound from any host star \citep{mroz2017no,gould2022free,sumi2023free}. Correcting for survey biases, these discoveries indicate that such ``free-floating planets'' (FFPs)---also commonly referred to as ``rogue planets''---are common, at of order $\approx20$ FFPs with masses $0.33<M/M_\oplus<6660$ per star in the Milky Way \citep{sumi2023free}. 

While planetary ejection is a ubiquitous process, the specific outcomes of this process are elusive to directly pinpoint. Ejected planets have been removed from their host systems such that they are not, by default, illuminated by any bright source. This renders them effectively invisible to the historically lucrative transit and radial velocity methods that have been leveraged to discover and characterize most known exoplanets to date \citep{christiansen2025nasa}. Furthermore, some planetary ejection mechanisms suggest that FFPs may stem from birth populations that are largely distinct from the well-characterized census of bound exoplanets---a near-inevitability, for example, for many planets born into orbits that are not long-term stable.

This poses a challenge as well as an opportunity. Planetary ejection studies offer a complementary avenue toward understanding regions of parameter space in planetary systems that are only minimally accessible with most exoplanet detection and characterization methods. A bridge between the theoretical and observational constraints on planetary ejection provides a crucial piece of the puzzle when interpreting the demographic-level evolutionary cycles of planetary systems.

\begin{figure*}
    \centering
    \includegraphics[width=1.0\textwidth]{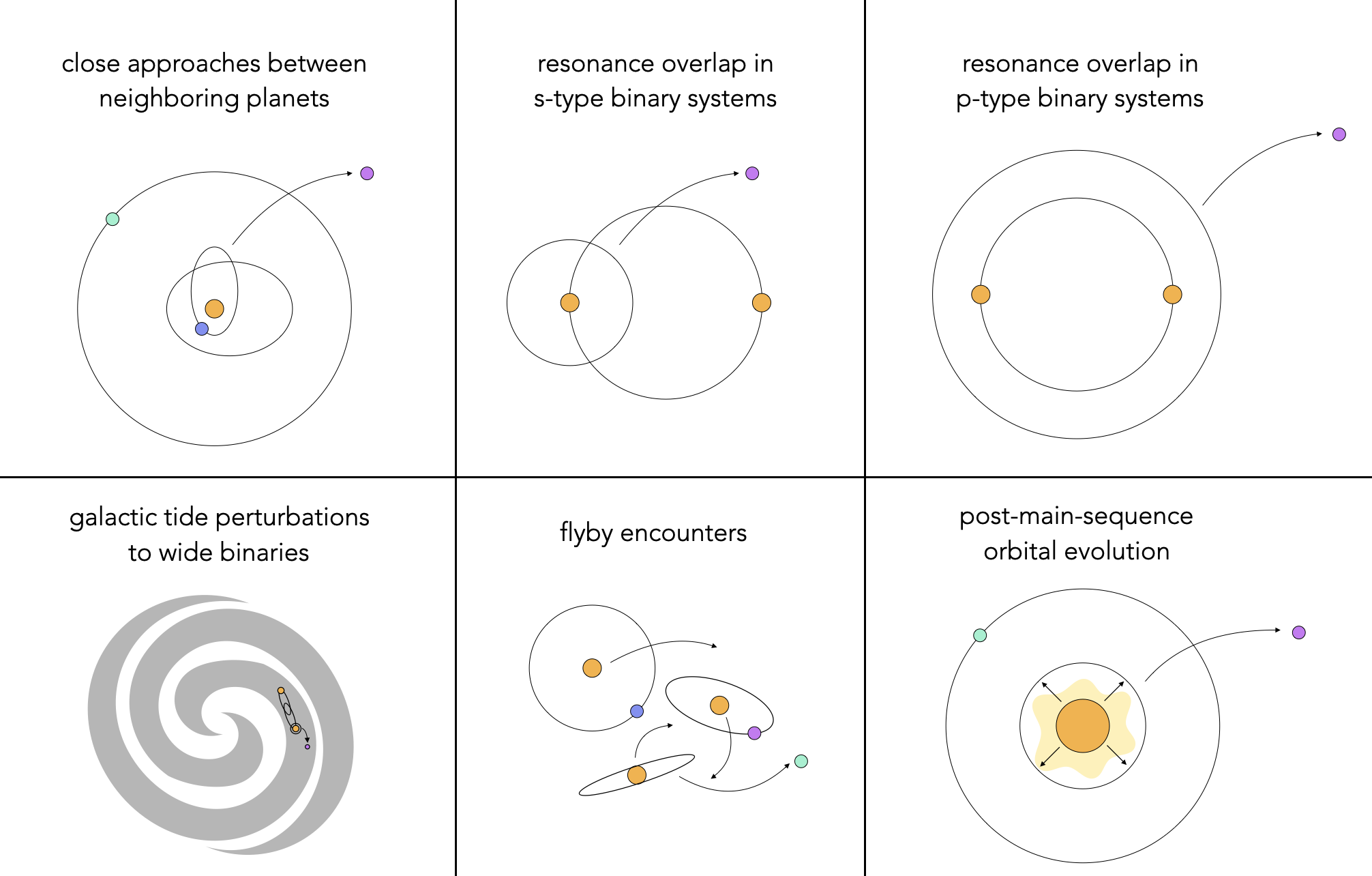}
    \caption{Schematic overview of instability-inducing mechanisms through which planets may be ejected from a system. These mechanisms, each discussed throughout this work, include close approaches between neighboring planets (top left; Section \ref{section:close_approaches}), binary-induced instabilities (top center and right, and bottom left; Section \ref{section:bound_stellar_companions}), flyby encounters (bottom center; Section \ref{section:flybys}), and post-main-sequence orbital evolution (bottom right; Section \ref{section:post-ms}).}
    \label{fig:schematic_ejection_mechanisms}
\end{figure*}

Depending on the birth properties and environmental factors shaping a planetary system, a range of dynamical mechanisms may induce instabilities that lead to planetary ejection. Indeed, any close encounters or instability-inducing mechanisms, such as planet-planet close approaches, dynamical perturbations from companion stars, stellar and planetary flybys, and host star evolution, have the potential to trigger the ejection of one or more initially bound planets. 

We provide a schematic overview of these instability-triggering mechanisms in Figure \ref{fig:schematic_ejection_mechanisms} for guidance. In this manuscript, which is designed as a pedagogical overview of the dynamics of planetary ejection, we begin by discussing the underlying mechanisms that may trigger planetary ejection. We first consider interactions between neighboring planets in Section \ref{section:close_approaches}, then examine the influence of bound stellar companions in Section \ref{section:bound_stellar_companions}. We describe potential ejection from flybys of unbound objects in Section \ref{section:flybys}. Lastly, we discuss the prospects for planetary ejection from post-main-sequence stellar evolution in Section \ref{section:post-ms}. We discuss the relative contributions of each planetary ejection mechanism as well as future prospects to constrain the planetary ejection rate in Section \ref{section:discussion}. We conclude with a summary of major takeaways in Section \ref{section:summary}.

\section{Close approaches between neighboring planets}
\label{section:close_approaches}

Stellar systems that host two or more planets around a single star often undergo instabilities in which neighboring planets experience close encounters. In this process, commonly referred to as ``planet-planet scattering,'' the close encounters may cause one or more planets to be scattered outward and unbound from the system. Conservation of the system's angular momentum implies that, in the process of planet-planet scattering, the lowest-mass planets tend to be ejected, whereas the highest-mass planets typically remain bound to the system. The remaining, bound planets are often left on dynamically excited (eccentric and/or inclined) orbits from the encounter.

In Section \ref{subsection:underlying}, we first outline some underlying principles that can be applied to describe systems' capacity for planetary ejection through planet-planet scattering. We then provide a brief overview of the evidence for planet-planet scattering in the solar system in Section \ref{subsection:pp_solarsystem} and an analogous overview for exoplanet systems in Section \ref{subsection:pp_exosystem}. We close with predictions for the free-floating planet population, drawn from single-star scattering scenarios, in Section \ref{subsection:pp_predictions}.

\subsection{Underlying principles}
\label{subsection:underlying}
In this section we describe quantitative metrics that provide intuition for planetary systems' typical proclivity toward planetary ejection, as well as relevant timescales for instability. 

\subsubsection{The Safronov number}
\label{subsubsection:safronov}
Depending on the physical conditions of the system, close encounters between two bodies in a planetary system may result in either a collision or an ejection. The outcome of a close encounter can be understood from first principles through examining a simple scenario in which two bodies undergo a grazing encounter.

\begin{figure}
    \centering
    \includegraphics[width=0.45\textwidth]{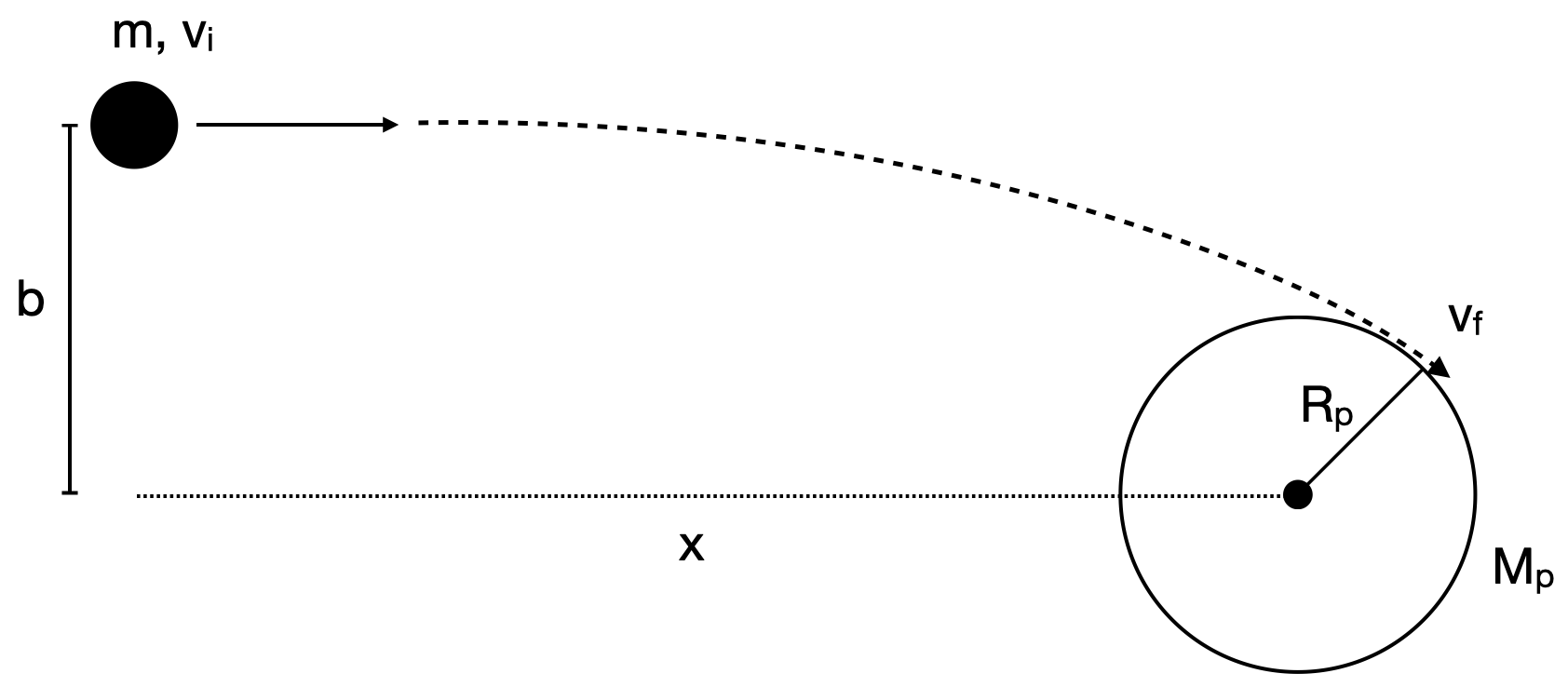}
    \caption{Schematic of the grazing-encounter scenario considered in Section \ref{subsubsection:safronov}, through which the Safronov number, which is used to determine the likelihood of collisions vs. ejections as the outcome of close encounters, is derived. Here $x$ is an initially large distance from which mass $m$, at initial velocity $v_i$, approaches a planet with mass $M_p$ and $R_p$ with impact parameter $b$. The final velocity $v_f$ of mass $m$ is denoted at the point of closest approach during its grazing encounter with mass $M_p$.}
    \label{fig:grav_focusing}
\end{figure}

Consider a body with mass $m$ approaching a planet with mass $M_p$ and radius $R_p$ from a large distance $x$, with impact parameter $b$ and initial velocity $v_i$ (Figure \ref{fig:grav_focusing}). Both $m$ and $M_p$ are in orbit around a star with mass $M_*$, where $M_*\gg m, M_p$. 

From conservation of energy, the final velocity $v_f$ when mass $m$ reaches a grazing encounter with mass $M_p$ can be obtained through the expression

\begin{equation}
\frac{1}{2}mv_i^2 - \frac{GM_p m}{\sqrt{x^2 + b^2}} = \frac{1}{2}{mv_f^2} - \frac{GM_p m}{R_p}.
\label{eq:full_energy_conservation}
\end{equation}
For $x\rightarrow\infty$, $1/\sqrt{x^2 + b^2}\approx0$ and Equation \ref{eq:full_energy_conservation} simplifies to 

\begin{equation}
v_f^2 = v_i^2 + v_{\rm esc,p}^2,
\label{eq:v_f}
\end{equation} 
where the escape velocity $v_{\rm esc, p}$ from planet $M_p$ is given as

\begin{equation}
    v_{\rm esc,p} = \sqrt{\frac{2 G M_p}{R_p}}.
\label{eq:escape_velocity_planet}
\end{equation}
To examine the case of ejection, we are interested in the scenario in which mass $m$ reaches a final velocity $v_f$ above the escape velocity of the system $v_{\rm esc,*}$. That is, we require 

\begin{equation}
v_f>v_{\rm esc,*} \,\, \mathrm{(for\, ejection)},
\label{eq:v_f_gtr}
\end{equation}
where

\begin{equation}
    v_{\rm esc,*} = \sqrt{\frac{2 GM_*}{r}}
    \label{eq:escape_velocity_star}
\end{equation}
to escape from an orbit around the host star, for planet-star distance $r$ at a given time.

Inspection of Equation \ref{eq:v_f} together with the condition of Equation \ref{eq:v_f_gtr} reveals that for any encounter velocity $v_i$, close encounters with a neighboring planet will lead to a boost above the system's escape velocity for $v_{\rm esc,p}^2/v_{\rm esc, *}^2 > 1$. Thus, the parameter $\Theta$, defined as

\begin{equation}
\Theta \equiv \frac{v_{\rm esc,p}^2}{v_{\rm esc, *}^2},
\label{eq:theta}
\end{equation}
is useful in demonstrating whether encounters will typically lead to ejections---the alternative being either (1) collisions or (2) close approaches in which both bodies remain bound to the host star without merging. $\Theta$ is commonly referred to as the ``Safronov number,'' and, in this simplified scenario, ejections occur where $\Theta>1$. Rearranging using Equations \ref{eq:escape_velocity_planet} and \ref{eq:escape_velocity_star}, the Safronov number may also be written as
\begin{equation}
\Theta = \frac{M_p r}{R_p M_*}.
\label{eq:theta_ecc}
\end{equation}
Equations \ref{eq:theta} and \ref{eq:theta_ecc} provide a generalized version of the Safronov number, which is often approximated specifically for circular orbits. For circular orbits, $a=r$ and Equation \ref{eq:theta_ecc} may be updated accordingly. In particular, the instantaneous orbital speed at a given separation $r$ from the star is

\begin{equation}
v_{\rm orb} = \sqrt{GM_*\Big(\frac{2}{r} - \frac{1}{a}\Big)}.
\label{eq:vorb}
\end{equation}
Therefore, in the limit of a circular orbit, $v_{\rm esc, *} = \sqrt{2}v_{\rm orb}$. In this case, the Safronov number may be rewritten as its commonly reported form
\begin{equation}
\Theta = \frac{1}{2}\frac{v_{\rm esc,p}^2}{v_{\rm orb}^2}\, \,\, (\mathrm{for\,} e=0).
\end{equation}

\begin{figure*}
    \centering
    \includegraphics[width=1.0\textwidth]{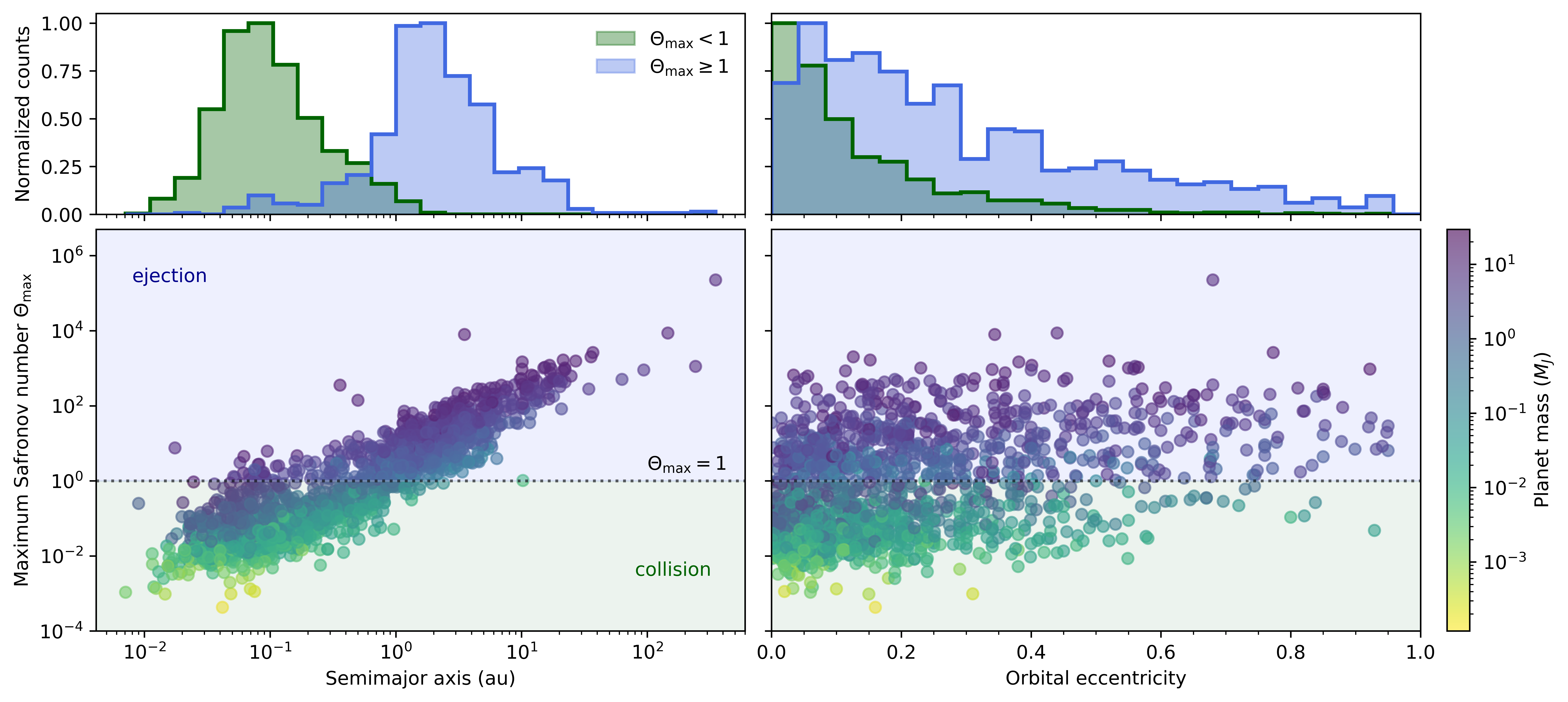}
    \caption{Maximum Safronov number of known, bound planets shown as a function of semimajor axis (left) and eccentricity (right), as a scatter plot (bottom) and histogram (top). The histograms are each normalized to peak at 1 for a clearer comparison between samples. Only planets with measured, nonzero uncertainties in orbital eccentricity are included. Though these planets remain bound to their host systems, the elevated eccentricities of planets with high Safronov numbers suggests that they may have ejected neighboring companions in their past histories. Because the Safronov number varies with location in the orbit, we take the maximum Safronov number $\Theta_{\rm max}$ corresponding to apoastron, where $r=a(1+e)$. The present-day maximum Safronov number calculated here is a proxy for the true value of interest, which is the pre-scattering Safronov number. This is an updated version of an earlier result first demonstrated by \citet{ford2008origins}.}
    \label{fig:safronov}
\end{figure*}

The above approximations assume no energy transfer between the planet and the ejected body, and they do not take into consideration the encounter geometry (instead assuming a grazing geometry), which in practice influences whether $m$ reaches the system's escape velocity \citep[e.g.][]{weidenschilling1975close}. Because of these limitations, the requirement for a planet to efficiently eject its neighbors may be more safely written as $\Theta\gg 1$. For planets in the regime $\Theta \ll 1$, close approaches by neighboring bodies tend to instead lead to collisions. 

Equation \ref{eq:theta_ecc} demonstrates that, in general, planets on wider orbits are more efficient at ejecting neighboring material. The majority of the currently known exoplanets were detected via the transit method, such that the census is heavily biased toward systems in which planets are difficult to eject: those in which planets are on extremely tight orbits with $\Theta\ll 1$, with a strong potential well that typically culminates in collisions, rather than ejections, from close encounters. \citet{wyatt2017how} developed a mapping of the typical outcomes from close approaches with the known census of exoplanets---assuming circular orbits and encounters with a single planet---demonstrating this bias.

Wider-orbiting known exoplanets---largely discovered through the radial-velocity or direct-imaging methods---often have higher Safronov numbers. \citet{ford2008origins} showed that planets with higher Safronov numbers typically have higher-eccentricity orbits than those with lower Safronov numbers, suggesting that neighboring planets may have been previously ejected from the systems. Updating this result with the current set of known planets (as of 7/1/2026) and using Equation \ref{eq:theta_ecc} to calculate the maximum Safronov number for each orbit, we demonstrate in Figure \ref{fig:safronov} that this finding continues to hold true for the present-day census of confirmed exoplanets.

\begin{figure*}
    \centering
    \includegraphics[width=1.0\textwidth]{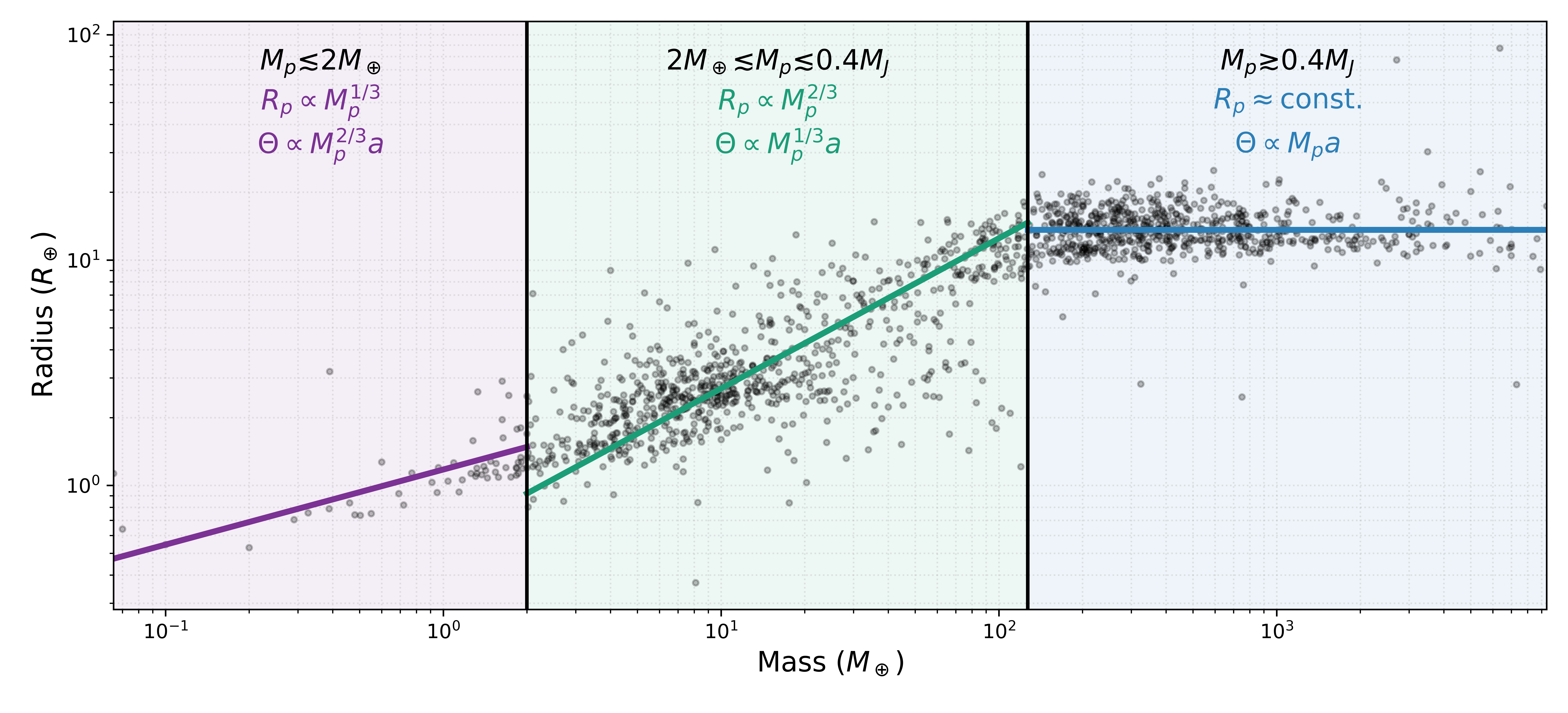}
    \caption{Theoretical, piecewise mass-radius scalings across planetary mass regimes, alongside associated scalings of the Safronov number. The census of observed exoplanets with masses and radii (including uncertainties) reported in the NASA Exoplanet Archive PSCompPars table as of 7/1/2026 is shown in gray, and the theoretical fits with fixed exponents are scaled to match this population. We note that the empirically-derived exponents fitted to each subpopulation in previous work have been observed to deviate slightly from these theoretical values, which enables improved matches at the boundaries between each subpopulation \citep[e.g.][]{chen2017radius}.}
    \label{fig:safronov_scalings}
\end{figure*}

Equation \ref{eq:theta_ecc} provides further intuition when considered in conjunction with planetary mass-radius relationships, as visualized in Figure \ref{fig:safronov_scalings}. For sub-Jovian-mass planets, the scaling of $R_p$ with $M_p$ is sub-linear, at roughly $R_p\propto M_p^{1/3}$ for $M_p\lesssim2M_{\oplus}$ and $R_p\propto M_p^{2/3}$ for $2M_{\oplus}\lesssim M_p\lesssim 0.4M_J$ \citep{valencia2006internal,chen2017radius}. In this regime, $M_p$ grows more rapidly than $R_p$. For a given stellar type and assuming $e=0$, this implies that $\Theta\propto M_p^{2/3}a$ for $M_p\lesssim2M_{\oplus}$ and $\Theta\propto M_p^{1/3}a$ for $2M_{\oplus}\lesssim M_p\lesssim 0.4M_J$. Hence, higher-mass rocky and Neptunian planets have higher Safronov numbers, such that more massive planets are generally more efficient at ejecting material for a given orbit and host star type.

For Jovian-mass planets ($M\gtrsim0.4M_J$), the mass-radius relationship is relatively flat at $R_p\propto M_p^0\approx\mathrm{constant}$ \citep{chen2017radius}. In this regime, given a known stellar type (and thus mass $M_*$), the Safronov number simplifies to scale as $\Theta\propto M_p r$, or $\Theta\propto M_p a$ for circular orbits. Thus, the mass dependence of the Safronov number is strongest for the highest-mass planets.

The Safronov number is not necessarily static for a given planet. A planet may migrate during or after its formation, either increasing or decreasing the Safronov number depending on the direction of migration. As giant planets cool and contract, altering their densities by a factor of a few, their Safronov numbers also accordingly shift, as explored by \citet{bonomo2017deeper} in the context of the CoRoT-9 b Jovian planet. This affects the branching ratios of accretion vs. ejection outcomes for close encounters.

The Safronov number also has important implications for the formation of planets. The concept of the Safronov number was originally developed in the context of gravitational focusing (see Section \ref{subsection:cross_sections} for further discussion of this mechanism), as a limit for planetesimal accretion to proceed \citep{safronov1969relative}. Thus, the fact that planets exist in our solar system with $\Theta>1$ implies that they must have either migrated or formed through a separate mechanism (in the case of the solar system giant planets, primarily gas accretion). This also implies a limit for planet formation: that is, the Safronov number may be used to approximate the widest orbital separation at which a planet of a given mass and radius, around a given host star, may form through planetesimal accretion such that $\Theta<1$ (accretion dominates over ejection). For an Earth-like planet around a solar-mass star, this limit is 14.2 AU. 

\subsubsection{Angular momentum deficit}
\label{subsubsection:amd}

The angular momentum deficit (AMD) is defined as the difference in angular momentum between a planetary system and the same system in which all planets are on circular, aligned orbits at the same semimajor axes. The AMD may be calculated as \citep{Laplace1878OeuvresI,laskar1997large, laskar2000on}

\begin{equation}
\mathrm{AMD} =\sum^{N}_{k=1}M_{p,k}\sqrt{GM_* a_k}(1 - \sqrt{1 - e_k^2}\cos i_k)
\end{equation}
where $M_{p,k}$, $a_k$, $e_k$, and $i_k$ are the mass, semimajor axis, eccentricity, and inclination of the $k$th planet, and the sum is taken over $N$ bound planets in the system. In this expression, the inclinations are taken relative to the invariable plane of the planetary system \citep{Laplace1878OeuvresI}, which is defined as the plane perpendicular to the net angular momentum vector of the system.

For sufficiently high AMD, secular evolution can be chaotic and the planets' orbital eccentricities and inclinations may diffusively evolve toward equipartition of the secular degrees of freedom, in some cases resulting in chaotic evolution and orbit crossings \citep{lithwick2011theory, wu2011secular, lithwick2014secular}. \citet{laskar2017amd} presents an analytic limit for the critical AMD required for stability in compact multi-planet systems, above which the system can undergo collisions and ejections commonly driven by overlapping mean-motion resonances \citep{chirikov1979universal,quillen2011three,petit2020path,tamayo2021criterion}. The innermost planet is typically most susceptible to significant AMD transfer \citep{wu2011secular}, which raises its eccentricity and in some cases induces instabilities that remove the inner planet from the system. In this process, the eccentricities and inclinations of the outer planets are pushed toward lower values to conserve angular momentum, such that the loss of the innermost planet culminates in a reduced system AMD.

\citet{hadden_free_2025} applied numerical simulations to demonstrate that detached planets reach their final state when there is no longer sufficient AMD to induce orbit crossings over time. Lower-mass planets in crowded systems undergo instabilities over longer timescales than higher-mass planets \citep{chambers1996stability, adams2003migration, marzari2025planet}. As a result, \citet{hadden_free_2025} found that while systems of scattered, detached equal-mass giant planets ($\gtrsim10M_{\rm Nep}$) often reach an AMD-stable configuration within 100 Myr after the first orbit crossing, comparable systems of lower-mass Neptune analogues can take Gyr to reach this point.

\subsection{Evidence for scattering-induced planetary ejection in the solar system}
\label{subsection:pp_solarsystem}

Planet-planet scattering and subsequent planetary ejection have been invoked to explain many observables in planetary systems, including the solar system. Here we briefly discuss the evidence for planetary ejection in the solar system, which serves as a relatively well-constrained case study. 

Solar system models have long incorporated the influence of planetesimal-driven migration, through which planetesimals are ejected from the system as the giant planets migrate \citep[e.g.][]{fernandez1984some, malhotra1993origin, malhotra1995origin, thommes1999formation, levison2001could,gomes2005origin,tsiganis2005,morbidelli2005chaotic, batygin2010early, levison2011late}. Over time, these models have evolved to consider possible additional planets that may have existed in the outer solar system during its past.

Several early works examined or alluded to the possibility that the solar system may have once included additional planets that no longer remain bound. For example, numerical simulations by \citet{thommes1999formation,thommes2002formation}, which were focused on modeling the formation of Uranus and Neptune, included scenarios in which a fifth giant planet formed and was subsequently ejected from the outer solar system. \citet{goldreich2004planet}, too, made the claim that some planet-sized objects were likely ejected from the outer solar system at the late stages of planet formation. If the outer solar system planets formed in a packed oligarchy, then up to five Neptune-sized isolation masses could have fit between 15-25 AU at mutual separations of 5 Hill radii---a configuration that would have destabilized as the natal disk dispersed, often resulting in the ejection of one or more planets \citep{ford_formation_2007}. 

\citet{nesvorny_young_2011} and \citet{batygin2012instability} each demonstrated that numerical simulations invoking a fifth solar system giant planet---which is often ejected\footnote{In a small fraction of runs reported by \citet{batygin2012instability}, an ice giant instead merged with Saturn, similarly reproducing a configuration comparable to that of the solar system. Thus, the five-giant-planet solar system model does not require that planetary ejection is invoked to reproduce present-day outcomes.} during the solar system's instability-driven evolution---provide an excellent match to solar system observables, such as the solar system's secular modes. Follow-up work by \citet{nesvorny_statistical_2012} examined a range of possible resonant orbital configurations and showed that the inclusion of a fifth (and possibly sixth) planet in the numerical simulations enables models to much more consistently match $a$, $e$, and $i$ of the solar system planets to present-day values.

Further numerical studies adopting this five-giant-planet+orbital instability model have demonstrated that it appropriately reproduces the capture of the Trojan asteroids \citep{nesvorny2013capture} and irregular satellites \citep{nesvorny2014capture} in the Jupiter system; that it is possible to maintain the Galilean moon system around Jupiter despite close-approach encounters between Jupiter and the extra ice giant \citep{deienno2014orbital,cloutier2015could}; and that it may enable inclination excitation for Iapetus relative to Saturn's orbital plane \citep{nesvorny2014excitation}, but that ejection of the fifth giant by Jupiter (rather than Saturn) is favored to avoid disrupting Iapetus's orbit \citep{cloutier2015could}. A fifth giant planet may also support the secular-resonance-driven dispersal of primordial asteroid families \citep{brasil2016dynamical}, and it is consistent with the predicted orbital evolution of Neptune necessary to reproduce the observed structure of the Kuiper belt \citep{nesvorny2015evidence, nesvorny2015jumping, deienno2017constraining}. Studies examining other aspects of the solar system's primordial configuration now often incorporate this fifth ice giant \citep[e.g.][]{clement2018mars, clement2021bornextra, clement2021born,liu2022early, kaib2024more,raymond2026was}.

Whether or not a fifth ice giant was ejected in the early solar system's history, it would be surprising if lower-mass planets were not. Collisions between planetary embryos have been modeled to build the cores of the solar system giant planets, and in this process a subset of the embryos (typically ranging from $\approx1-6\,M_{\oplus}$) is commonly ejected and removed from the system \citep[e.g.][]{jakubik2012accretion, izidoro2015accretion}. Additional planetary embryos may also be ejected at other points in the solar system's evolution---for example, in simulations presented within \citet{clement2018mars}, multiple embryos from $0.25-2.5\,M_{\rm Mars}$ were removed from the inner solar system during the giant planet instability. If planetary embryos remained in the outer solar system at later times, after gas giant formation and disk dispersal, most would have been similarly ejected \citep{barclay2017demographics,silsbee2018producing}. Though the true primordial occurrence and size distribution of such lower-mass ejected planets in the solar system's early history remains relatively poorly constrained, the pervasiveness of planetary ejection in a bottom-up planet formation framework strongly suggests that it has played a role in the solar system's early evolution. 

\subsection{Evidence for scattering-induced planetary ejection in extrasolar systems}
\label{subsection:pp_exosystem}

Beyond the solar system, hints in the architectures of extrasolar systems also suggest a high rate of planet-planet scattering. Such evidence is typically drawn from the spacings and orbital properties of planetary systems, which often point toward dynamical sculpting beyond the planets' initial formation. 


Perhaps the most direct line of evidence for a high rate of planetary ejection in extrasolar systems is the eccentricity distribution observed for giant exoplanets. Planet-planet scattering has long been invoked to explain the elevated eccentricities of exoplanet orbits, often in the context of hot Jupiter formation and high-eccentricity giant exoplanet orbital configurations \citep{rasio1996dynamical, weidenschilling1996gravitational,lin1997on,adams2003migration,chatterjee2008dynamical,nagasawa2008formation,raymond2009planet}. Even extremely high eccentricities $e>0.99$ may be reached by planet-planet scattering \citep{carrera2019planet}.

Population-level studies have found that the high eccentricities of giant exoplanet systems are well-matched by models in which nearly all ($>75-80\%$) systems underwent an orbital instability \citep{juric2008dynamical,raymond2010planet,raymond2011debris}. Planets with a higher Safronov number also show a wider spread in orbital eccentricities \citep[][see also Figure \ref{fig:safronov} in this review]{ford2008origins}, suggesting that at least a subset of these planets may have been dynamically excited to higher eccentricities through past scattering events. Planet-planet scattering notably offers a pathway toward elevated eccentricities while simultaneously reproducing observed features of the spin-orbit distribution for giant exoplanet systems \citep{rice2022origins, rusznak2025from, esposito2026unified, dong2026planet}.

Another line of evidence used to argue for a high rate of orbital instabilities in exoplanet systems is the relative spacings of observed planets. The dynamical separation of planets may be measured with the critical spacing $\Delta$, given as 

\begin{equation}
\Delta\equiv \frac{a_2 - a_1}{R_H}
\end{equation}
for semimajor axes $a_1$ and $a_2$, and Hill radius 

\begin{equation}
R_H = \frac{a_1 + a_2}{2}\Big(\frac{m_1+m_2}{3M_*}\Big)^{1/3}
\end{equation}
with planet masses $m_1$ and $m_2$. \citet{gladman1993dynamics} showed that systems of two planets on circular, coplanar orbits are stable to orbit-crossing instabilities for $\Delta>2\sqrt{3}\approx3.46$. For higher-multiplicity systems, this spacing must be larger to ensure stability, with specific configurations (planet multiplicities and mass ratios) examined in further detail in a slew of later studies \citep{chambers1996stability,zhou2007post,smith2009orbital, funk2010stability,pu2015spacing,tamayo2015dynamical,morrison2016orbital,obertas2017stability,rice2018survival,yalinewich2020nekhoroshev,bartram2021orbital,lissauer2021orbital,rice2023stable,volk2024differences,wu2025enhanced,outland2026orbital,gavino2026orbital}.

Early work by \citet{barnes2004in}, examining the 13 multiplanet systems known at the time, demonstrated that, while the known systems were stable, many fell precariously close to the edge of orbital instability. This led to the ``packed planetary systems'' hypothesis \citep{barnes2004predicting}, which suggests that \textit{all} multiplanet systems are tightly packed, and those that appear to have larger spacings instead host planets below the detection limit within the observed gaps. This tight dynamical packing was later found to extend to many planetary systems that lack nearby gas giant planets \citep{lovis2011harps}, including at least $\approx30\%$ of multiplanet systems detected through the \textit{Kepler} space mission \citep{borucki2010kepler}, pointed out by \citet{fang2013are}.

Building upon these findings, \citet{pu2015spacing} showed that \textit{Kepler} planets' spacings lie just beyond the stability threshold for survival over Gyr timescales, at $\Delta\approx12$, leading the authors to propose that most short-period planetary systems originated in dynamically packed configurations that subsequently underwent instabilities. The instabilities in these systems would largely result in collisions due to the short-period orbits of \textit{Kepler} systems, such that they would not directly produce ejections. However, this finding does point toward a high incidence of dynamical sculpting within planetary systems---where analogous processes at wider orbits would result in ejections rather than collisions.

The paucity of mean-motion resonances (that is, orbital commensurabilities---formally, planet pairs with at least one librating resonant angle) in exoplanet systems, too, has been called upon as evidence for a high rate of orbital instabilities. \citet{lissauer2011architecture} found that \textit{Kepler} planetary systems reported from the first four months of the mission \citep{borucki2011characteristics} show little to no statistical preference toward mean-motion resonance, with most systems lying wide of dominant resonances by up to a few percent. This result was further affirmed by \citet{fabrycky2014architecture} using a larger set of planet candidate from 16 months of \textit{Kepler} data \citep{batalha2013planetary}. 

The low occurrence of mean-motion resonances in exoplanet systems is unexpected as a direct consequence of formation. Commensurabilities are predicted to be common in planetary systems: the Jovian and Saturnian moon systems show an overabundance of mean-motion resonances \citep{roy1954on}, and resonances play a key role in shaping the solar system's structure \citep{peale1976,malhotra1998orbital}. Planetary migration models oriented toward exoplanet systems predict that planets should typically be found at or near commensurability after migration within the protoplanetary disk \citep{lee2002dynamics, cresswell2006on,terquem2007migration,raymond2008observable, ogihara2009n}. 

The low rate of present-day mean-motion resonances in the census of mature planetary systems implies a high rate of orbital instabilities during or soon after protoplanetary gas disk dispersal \citep[e.g.][]{ogihara2009n, izidoro2017breaking,izidoro2021formation,liu2022early}---or, as required to reproduce the suggestive high rate of mean-motion resonances in young ($\approx10-60$ Myr) planetary systems \citep{dai2024prevalence}, at delayed stages post-disk-dispersal. Suggested mechanisms for later-stage resonant chain disruption include delayed instability caused by remnant planetesimal disks in the systems \citep{griveaud2024solar,lorusso2026planetesimal,choksi2026two}. While the low rate of commensurabilities in \textit{Kepler} systems is specific to transiting planets, which lie in a parameter space in which instabilities would generally result in collisions rather than ejections, they may be indicative of system-wide disruption corresponding to planetary ejection in the outer system \citep[e.g.][]{goldreich2004final}.


Case studies of individual exoplanet systems have provided evidence that planet-planet scattering and consequent ejection may have occurred in the systems' histories, as well. \citet{ford2005planet}, for example, demonstrate that the excited orbital eccentricities in the 3-planet Upsilon Andromedae system may be naturally reproduced by a brief, $\approx1000$-year period of past instability and the ejection of a fourth planet. \citet{bonomo2017deeper} carried out a case study of the CoRoT-9 planetary system, which contains a single eccentric warm Jupiter planet on a $\approx$95-day orbit, to show that the previous ejection of a second, $\approx50M_{\oplus}$ planet well reproduces the system architecture. \citet{yuan2024scattering} and \citet{lu2026architecture} present case studies for the 2-planet GJ 1148 and 14 Herculis systems, respectively, showing that if the high eccentricities of the systems result from planet-planet scattering, then a third planet was most likely ejected from the systems.

\subsection{Predictions and implications: planet-planet scattering}
\label{subsection:pp_predictions}

\subsubsection{The present-day population of bound planets}
\label{subsubsection:pp_clues_bound}
FFPs ejected by planet-planet scattering typically leave behind equal- or higher-mass planets that remain bound to the system. As a result, limits may be placed on the production of FFPs from planet-planet scattering by examining the occurrence of high-mass bound exoplanets that show possible signatures of past instabilities.

The orbital configurations of giant exoplanet orbits suggest that instabilities are common, with an instability occurrence rate $\gtrsim75-80$\% that well reproduces the eccentricity distribution of Jupiter-mass planets \citep{juric2008dynamical, raymond2010planet,raymond2011debris}. In the limit that \textit{all} FFPs are ejected from their natal systems through planet-planet scattering with giant-planet neighbors, the number of FFPs per star $N_{\rm free}/N_{\rm stars}$ may be calculated as
\begin{equation}
    \frac{N_{\rm free}}{N_{\rm stars}} = f_{\rm giant} f_{\rm unstable} n_{\rm ejec},
\end{equation}
where $f_{\rm giant}$ is the fraction of stars hosting giant planets (in particular, those with $\Theta\gg1$), $f_{\rm unstable}$ is the fraction of those systems that undergo instabilities, and $n_{\rm ejec}$ is the typical number of planets ejected from each system that undergoes an instability \citep{veras_planetplanet_2012,miret2022rich}. 

The occurrence of wide-orbiting Jovian-mass planets is now relatively well-constrained in FGK star systems through a combination of microlensing \citep{suzuki2016exoplanet}, direct imaging \citep{bowler2016imaging,baron2019constraints,nielsen2019gemini}, and radial velocity surveys \citep[][see also \citet{clanton2016synthesizing} for a synthesis of these methods]{cumming2008keck,mayor2011harps,fernandes2019hints,wittenmyer2020cool,fulton2021california}. \citet{miret2022rich} demonstrated that in principle (that is, from an ejection rate standpoint), up to 100\% of the highest-mass free-floating planets may be accounted for through single-star dynamical instabilities and subsequent ejection. Notably, this work focused primarily on high-mass young ``planets'' in star-forming regions, with formation pathways that remain under debate. 

We repeat this exercise here, adopting the microlensing-constrained FFP mass function from \citet{sumi2023free}

\begin{equation}
\frac{dN}{d\log_{10} M} = 2.18\left(\frac{M}{8\,M_\oplus}\right)^{-0.96}\;\mathrm{dex}^{-1}\,\mathrm{star}^{-1},
\end{equation}
which corresponds to 0.09 free-floating giant planets per star between 0.3 and 13 $M_J$. If all free-floating giant planets were produced via ejection by planet-planet scattering, then roughly one giant planet ($M>0.3M_J$) would need to be ejected per instability for a giant-planet occurrence rate $\approx10-15\%$ \citep{clanton2014synthesizing} and instability rate 75-80\%. The adopted giant-planet occurrence rate for this estimate is specific to M dwarf systems from \citet{clanton2014synthesizing}, in accordance with typical representatives of the galactic stellar population.

Jupiter-mass planets may not, however, be the sole (or even dominant) ejectors of their neighbors. Wide-orbiting Neptune-mass planets, too, may play a leading role. Observational limits from microlensing surveys \citep{suzuki2016exoplanet, poleski2021wide}, alongside indirect evidence from axisymmetric protoplanetary disk substructures at tens of AU host-star separations \citep{andrews2018disk,zhang2018disk}, indicate that these wide-orbiting Neptunes are common. Recent studies have demonstrated that wide-orbiting Neptunes may be the dominant ejectors in many systems, resulting in the removal of equal- or lower-mass neighboring planets \citep{hadden_free_2025,guo_formation_2025,lorusso2026planetesimal}. 

\subsubsection{The mass distribution of FFPs}

The initial conditions of planetary systems---in particular, their typical spacings, orbit distributions, and masses---play a key role in setting the expected outcomes of planet-planet scattering. We describe population-level simulated trends for the outcomes of planet-planet scattering in this section, noting that specific predictions are strongly dependent upon the assumed initial conditions of the systems.

In systems with unequal-mass planets, lower-mass planets tend to be most efficiently ejected through scattering---generally pushing the mass function of ejected planets to be more bottom-heavy. However, inner low-mass planets may, in some cases, eject more massive, longer-period planets on high-eccentricity orbits \citep{2026ApJ..1000..179Z}. Other competing effects further complicate the interpretation of the mass function. 

In equal-mass multiplanet systems, lower-mass ensembles and those with larger mutual separations typically take longer to eject planets \citep[e.g.][]{chambers1996stability, marzari2002eccentric,faber2007total,chatterjee2008dynamical,petit2020path}, with a timescale dependence on the planet mass and the innermost planet's orbital period \citep{hadden_free_2025,zhai2025dynamical}. For equal-mass, two-planet Hill stable planetary systems with $e\lesssim0.3$, \citet{veras2013simple} found a representative instability timescale

\begin{equation}
    \log_{10}N_{\rm orb}\approx5.2\Big(\frac{\mu}{M_J/M_{\odot}}\Big)^{-0.18}
\end{equation}
given a mass ratio $\mu$ between each planet and the host star, corresponding to a number of inner planet orbits $N_{\rm orb}$.
For a suite of simulations with five equal-mass planets and constant fractional spacing between planets, \citet{hadden_free_2025} similarly found an empirical relation

\begin{equation}
    T_{\rm ej}(M_p)\approx2.9\times 10^7 \Big(\frac{M_p}{M_{\rm Nep}}\Big)^{-1.64} P_{1,0}
\end{equation}
for the time taken for the number of bound planets to decrease by one, with initial orbital period $P_{1,0}$ and where $M_{\rm Nep}$ corresponds to one Neptune mass. Thus, lower-mass crowded systems take longer to undergo instabilities than higher-mass crowded systems.

Because of this difference in instability timescale with planet mass, populations produced from early- vs. late-time instabilities may have differing properties. A natural time for the first wave of instabilities and ejections is early in a system's lifetime, as the mass distribution substantially shifts around the era of planet formation and disk dispersal. These instabilities would generally occur within the first few Myr of a system's lifetime \citep{desousa2020dynamical}, with possible delays in the instability timescale to within the first few hundred Myr based on the observed longevity of resonant chains in inner planetary systems \citep{dai2024prevalence,griveaud2024solar,lorusso2026planetesimal,choksi2026two}. 

The first wave of instabilities should be bottom-heavy in its mass distribution, as the least massive planets would be ejected first. \citet{guo_formation_2025} (see also \citet{ma2016free}) leveraged population-synthesis models from \citet{ida2018slowing} to forward-model the production of FFPs from planet-planet scattering in single-star systems within the core accretion paradigm. The authors found that most planets are ejected during instabilities induced during these earlier stages, with a predicted mass distribution of FFPs that is highest (and roughly flat) around 0.1-10$M_{\oplus}$. Notably, higher-mass self-gravitating ``fragments'' of material---in some cases planet-mass---may be more readily ejected through early instabilities in the gravitational instability formation framework \citep{papaloizou2001orbital,forgan2015dynamical,forgan2018towards,schib2025dipsy,schib2025dipsyii}, with distinct implications from the core-accretion paradigm generally discussed within this section.

Over the remainder of the system's lifetime on the main sequence, additional instabilities may be triggered as well. These slower-acting instability mechanisms, induced by secular drift and AMD transfer, would most efficiently remove \textit{higher}-mass planets, as discussed in \citet{hadden_free_2025}. Specifically, high-mass planets in multi-giant-planet systems would be the first to be ejected through these mechanisms, whereas lower-mass sets of planets would be much longer-lived if they survived the initial phase of early instabilities.

Planets are not the only bodies that may become unbound through the process of scattering described throughout this section. Planet-planet scattering may also produce a large population of free-floating moons, though some FFPs with sufficiently short-period moons may retain their satellites even after ejection \citep{debes2007survival,gong2013effect, hong_innocent_2018, rabago2019survivability,huang2026free}. Tidally-induced migration in star-planet-moon systems \citep{alvarado2017effect}, too, has the potential to trigger scattering events that destabilize and eject moons, with the efficiency of this process dictated by the system geometry and the host planet's tidal quality factor and Love number \citep{tokadjian2020impact,sucerquia2020can,dobos2021survival,bolmont2025survival}.

Furthermore, planet-planet scattering among wide-orbiting giant planets has been shown to efficiently eject planetary embryos \citep[e.g.][]{chambers2001making,veras2005influence, raymond2005terrestrial, barclay2017demographics}, as well as planetesimals \citep[e.g.][]{raymond2010planet, raymond2011debris, raymond2012debris, marzari2014impact, raymond2018implications, rice2019hidden} that may contribute toward the high number density of free-floating interstellar objects inferred from recent detections \citep{do2018interstellar}. These bodies would be ejected at a range of overlapping timescales: for example, planetary embryos would generally be ejected early in a system's lifetime, but they may heavily overlap in their mass distribution with moons that are commonly ejected alongside giant planets.

\subsubsection{The velocity distribution of FFPs}
Another observable that offers insight into the relative prevalence of FFP production mechanisms is the distribution of FFP excess velocities  $v_{\infty}$. The excess velocity is defined as 

\begin{equation}
v_{\infty} = \sqrt{v^2 - v_{\rm esc,*}^2},
\end{equation}
where $v$ is the velocity of the ejected planet at the distance where the body becomes unbound, and the escape velocity $v_{\rm esc,*}$ from the host star at that same position may be drawn from Equation \ref{eq:escape_velocity_star}.

\citet{huang2026free} derived a useful expression for the critical velocity $v_{c,\rm 2pl}$ around which $v_\infty$ is peaked for two-body scattering, provided as 

\begin{equation}
v_{c,\rm 2pl}= \Big(\frac{Gm_1}{0.12 a_2}\Big)^{1/2}\Big(\frac{m_1}{m_1+m_2}\Big)\Big(\frac{a_1}{a_2}\Big)^{1/4}
\end{equation}
for two-planet systems with semimajor axes $a_1$, $a_2$ and planetary masses $m_1$, $m_2$. This expression was obtained by combining close-encounter-induced ejection rate estimates from \citet{pu2021strong} and \citet{li2022long} with principles of energy exchange. A comparable expression was also obtained by \citet{huang2026free} for the critical velocity $v_{c,\rm 3pl}$ in the case of 3-body scattering, given as 

\begin{equation}
    v_{c,\rm 3pl}= \Big(\frac{Gm_{\rm max}}{0.12 a_3}\Big)^{1/2}\Big(\frac{m_{\rm max}}{m_{\rm max}+m_{\rm ejec}}\Big)\Big(\frac{a_1}{a_3}\Big)^{1/4}.
\end{equation}
Here $m_{\rm max}$ is the highest-mass planet in the system, while $m_{\rm eject}$ is the mass of the ejected planet and $a_1$, $a_3$ are the semimajor axes of the innermost and outermost planet, respectively. In both expressions, the critical velocity corresponds to the peak of a probability distribution of velocities, which has a larger width for systems with more constituent bodies.

Ejected planets from planet-planet scattering are predicted to have relatively low excess velocities $<10$ km/s, with a median at $\approx2-3$ km/s and lower excess velocities attained for planets ejected from larger initial semimajor axes \citep{coleman_predicting_2025, bhaskar_properties_2025,zhai2025dynamical}. Ejected planet-mass, self-gravitating fragments formed from gravitational instability, too, have comparably low expected excess velocities $\approx4-5$ km/s \citep{forgan2018towards}. \citet{bhaskar_properties_2025} showed that planets are typically ejected within 30$\degree$ of the plane of the planetary system, such that their trajectories, in addition to their velocities, may encode some information about their birth environments.

\section{Instabilities in systems with bound stellar companions}
\label{section:bound_stellar_companions}

Binary and multiple-star systems are ubiquitous across the galaxy: roughly half of sun-like stars host a bound stellar companion, with a higher occurrence of companions to higher-mass stars \citep{duquennoy1991multiplicity, raghavan2010survey, duchene2013stellar, offner2023origin}. Planets in multi-star systems are susceptible to many of the same dynamical instability mechanisms that induce planetary ejection in single-star systems, such as planet-planet scattering and intra-system resonance overlap. In addition to these mechanisms, planets in binary systems may undergo further instabilities caused by perturbations from the companion star. 

Exoplanets in multiple-star systems may be divided into two classes: circumstellar (``$s-$type,'' historically standing for ``satellite-type'') and circumbinary (``$p-$type,'' historically standing for ``planet-type'') planets \citep{dvorak1984numerical}. Circumstellar planets (top center panel of Figure \ref{fig:schematic_ejection_mechanisms}) orbit a single star in a binary or higher-multiplicity stellar system, such that these systems operate similarly to their single-star counterparts but with additional perturbations imposed by one or more additional, high-mass companions. Circumbinary planets (top right panel of Figure \ref{fig:schematic_ejection_mechanisms}) instead orbit two stars\footnote{Note that we restrict this category to binary star systems, as over a dozen circumbinary planets have been found \citep[e.g.][]{doyle2011kepler, martin2019bebop, thornton2026detection}, whereas no planets have been confirmed encircling higher-multiplicity systems---though see \citet{smallwood2021gw} for discussion of a tentative circumtriple planetary system in the GW Orionis system.} and therefore undergo qualitatively different evolutionary trajectories. 

The three-body problem, including the circular restricted three-body problem that may be applied to planetary systems in certain cases---circular orbits and large mass ratios between the planet and stars---has an extensive history in astronomy. There have been correspondingly many studies of three-body systems' stability, spanning both analytic \citep[e.g.][]{szebehely1977analytical, szebehely1980stability} and numerical \citep[e.g.][]{henon1970stability, black1982simple, pendleton1983further} results. 

Many of these foundational studies impose stringent criteria on the orbit and mass-ratio regimes for their analyses, limiting their applicability to true systems. Therefore, we focus here on relatively general forms that account for orbital eccentricities and unequal stellar mass ratios, as are commonly observed in binary star systems \citep{moe2017mind, el2019discovery, hwang2022eccentricity}. 

We emphasize that instability is a necessary but not sufficient condition for planetary ejection. Many binary systems undergo instabilities that do not necessarily induce planetary ejection. Instabilities may instead lead to collisions with other planets or with stars within the system, or they may, in rare cases, introduce exchange reactions in which the planet becomes bound to a different star. Thus, we begin with an overview of instabilities in the context of potential planetary ejection, then discuss the implications of these instabilities for ejection rates.


\subsection{Instabilities of circumstellar planets in binary systems}
\label{subsection:s_type}

Of the possible configurations for circumstellar planets in non-single-star systems, circumstellar planets in binary star systems are the most well-studied to date. This is likely in part due to their relative simplicity, and in part because they are the most commonly observed configuration to date for non-single-star planetary systems. 

We divide instabilities of circumstellar planets in binary systems into two regimes: those most relevant for close binaries ($a\lesssim$ a few hundred au) where the instability limit is set by intra-system dynamics (Section \ref{subsubsection:close_binaries}), and those most relevant for very wide binaries ($a\gtrsim$ a few hundred au) where the instability limit is instead set by the influence of passing stars and the galactic tide (Section \ref{subsubsection:wide_binaries}). The instability limits described in the ``close stellar binaries'' overview may also play an important role in high-eccentricity wide binaries with close periastron approaches.

\subsubsection{Circumstellar planets in  close stellar binaries}
\label{subsubsection:close_binaries}
For circumstellar planetary orbits in close binary systems, there exists an outermost planetary semimajor axis for dynamical stability. Early derivations of this boundary were obtained through analytic approximations by \citet{szebehely1980stability} in the context of the planar, circular restricted three-body problem (enabling a wide range of binary mass ratios), as well as numerical calculations by \citet{rabl1988satellite} for circular planetary orbits, equal-mass stars, and coplanar, eccentric planet-binary configurations. \citet{benest1988stable,benest1988planetary, benest1993stable} extended these frameworks to examine stability in the elliptic planar restricted three-body problem, including case studies of known binary systems. 

\begin{figure}
    \centering
    \includegraphics[width=0.48\textwidth]{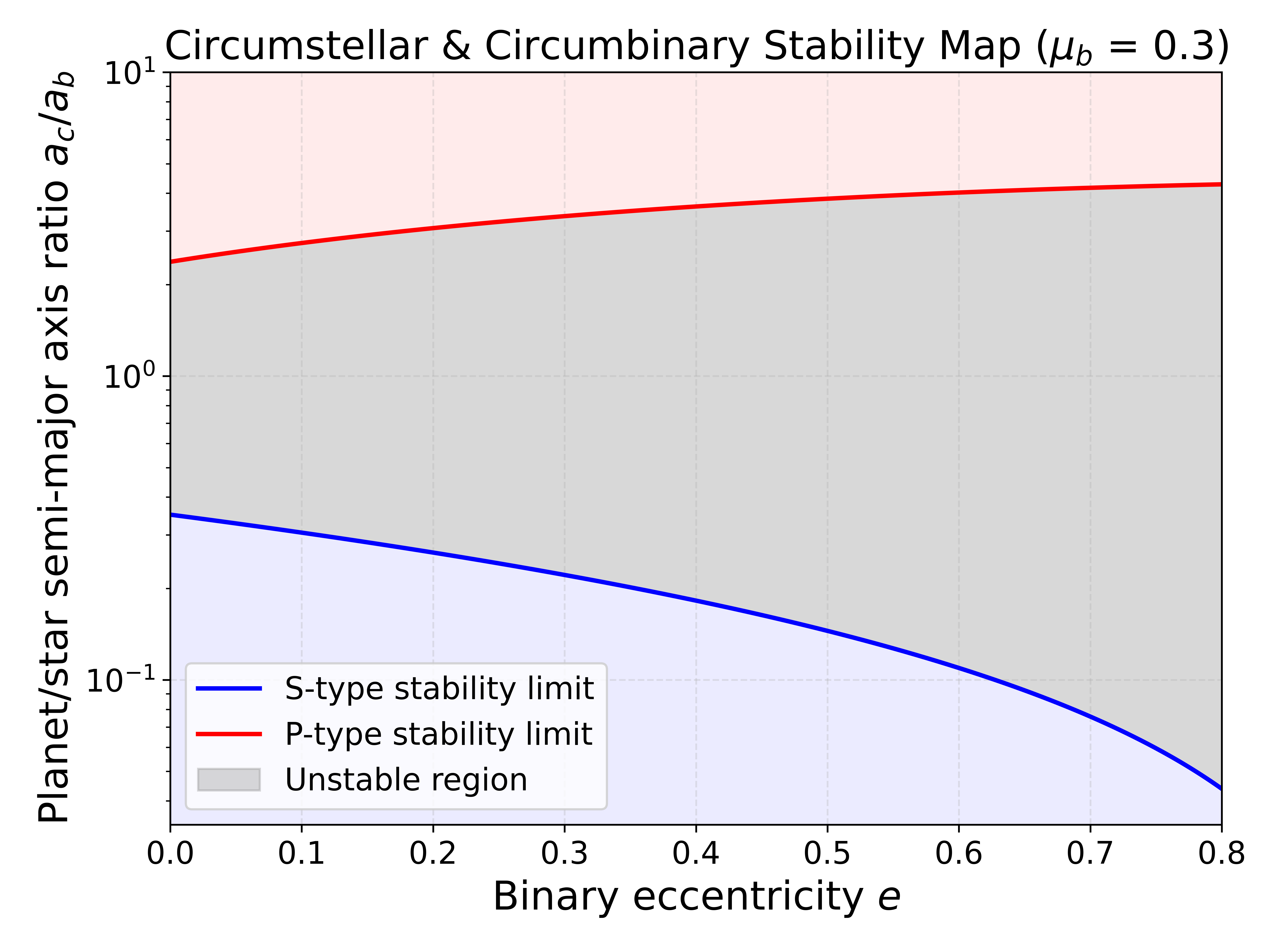}
    \caption{Example thresholds for $s-$type and $p-$type binary instabilities for stellar binary mass ratio $\mu_b=0.3$, displayed as a function of the binary eccentricity and the ratio between the planet and binary semimajor axis. Boundaries are drawn directly from \citet{holman1999long}, provided in this work as Equations \ref{eq:stype_stability} and \ref{eq:ptype_stability}.}
    \label{fig:stability_binary}
\end{figure}

\citet{holman1999long} later conducted a large suite of numerical simulations in the elliptic restricted three-body problem---that is, with eccentric binary orbits and massless, test-particle planets---spanning $10^4$ binary orbital periods and covering a range of binary separations and mass ratios. The commonly-adopted empirical formula for the outermost stable semimajor axis $a_c$ derived by \citet{holman1999long} is given as

\begin{equation}
\begin{split}
a_c/a_b = &(0.464 \pm 0.006) + (-0.380\pm0.010)\mu_b \\ 
&+ (-0.631\pm0.034)e + (0.586\pm0.061)\mu_b e \\ 
&+ (0.150\pm0.041)e^2 + (-0.198\pm0.074)\mu_b e^2
\end{split}
\label{eq:stype_stability}
\end{equation}
where $e$ is the binary eccentricity, $a_b$ is the binary semi-major axis, and $\mu_b$ is the binary mass ratio. This stability limit is shown in blue in Figure \ref{fig:stability_binary} for a canonical binary mass ratio $\mu_b=0.3$.

Notably, all planets were assumed to lie on circular orbits in \citet{holman1999long}. \citet{pilat2002stability} built upon this work to consider eccentric planet orbits using the Fast Lyapunov Indicator method \citep{froeschle1997fast}, which rapidly identifies chaotic, unstable orbits. With this method, the authors numerically examined the stability of circumstellar planetary systems over 1000 binary orbital periods, demonstrating that planet eccentricity plays a minor role in setting stability relative to binary eccentricity. More recently, computational advances enabled an extensive new suite of $\approx$700 million $N-$body simulations that extended the expressions of \citet{holman1999long} to non-coplanar orbits across the range of possible planetary orbital eccentricities \citep{quarles2020orbital}, considering integration times of $10^5$ yr per system. \citet{quarles2020orbital} provide a series of new coefficients to Equation \ref{eq:stype_stability} for varying binary eccentricity (Table 3 of \citet{quarles2020orbital}) and planetary inclination (Table 4 of \citet{quarles2020orbital}), with interpolation maps for intermediate values. 

Further studies have examined the stability of solar-system-like configurations in binary star systems, including a demonstration that a stellar companion supports planetary ejection during instabilities in the Jumping Jupiter model \citep{marzari2005jumping}, as well as a characterization of the planetary removal timescale within the instability region for Earth-like planets \citep{david2003dynamical}.\footnote{Note that the definition of ``ejection'' in \citet{david2003dynamical} encompasses both planets that become unbound from the host star and planets that collide with the host star; hence, to avoid confusion we use the term ``planetary removal'' in this context.} \citet{avila2026planetary}, too, carried out a suite of 2500 simulations following the 0.5-Gyr evolution of three giant planets on circumstellar orbits with a stellar binary companion at $a=50-1500$ AU, mapping the rate of planetary survivors as a function of binary properties. These contributions fold in the impact of neighboring planets and consider how planetary removal times scale with the periastron distance of the companion star. 

\citet{mudryk_resonance_2006} applied semi-analytic models in the coplanar regime to provide useful intuition for the physical mechanisms underlying these numerically-characterized instability boundaries. Specifically, they showed that secular forcing is strong for planetary systems with nearby stellar companions, leading to significant overlap between subresonances of individual mean-motion resonances that triggers chaotic diffusion and planetary instability. \citet{mudryk_resonance_2006} demonstrated that, in practice, this leads to rapid ejection dominated by the influence of low-order resonances, with a jagged instability boundary as a function of eccentricity for a given mass ratio. This result remains in rough agreement with boundaries derived through orbital integrations, but demonstrates the more complex structure of the true stability boundary.

Most of this section has focused on instabilities in systems with fully-formed stars and fully-formed planets. However, in the presence of a growing stellar perturber, recent simulations have suggested that both Jovian-mass \citep{calovic2026disc} and lower-mass \citep{nayakshin2026disc} planets formed through gravitational instability may be efficiently ejected through instabilities with the neighboring, growing stars. Such planets would be expected to have low excess velocities around 1-3 km/s. 

\subsubsection{Circumstellar planets in wide stellar binaries}
\label{subsubsection:wide_binaries}

Over secular timescales, wide binaries may be perturbed to high orbital eccentricities through gravitational interactions with passing stars and the galactic tide \citep{jiang2010evolution}, triggering orbital instabilities and planetary ejection. \citet{kaib2013planetary} simulated this scenario for systems hosting four solar-system-like giant planets and a wide, bound stellar companion at 1,000-30,000 AU separations, showing that $30-60\%$ of examined systems underwent at least one planetary ejection within 10 Gyr. \citet{correa2017galactic} extended this work to a broader set of wide binaries, excluding planets from their simulations but similarly finding that the widest (particularly $a>10,000$ au) binaries and those with the highest orbital eccentricities ($e>0.9$) have a high binary disruption rate that would potentially lead to corresponding planetary disruption.

Given the high rate of high-eccentricity wide binaries \citep{hwang2022eccentricity}, this mechanism may serve as an important reservoir for FFPs. Indeed, \citet{grishin2025hot} demonstrated that the ejection of giant planets from wide binary interactions with the galactic tide could account for $24^{+117}_{-15}$\% of the population of free-floating Jovian-mass planets. This mechanism may contribute a significant subset of FFPs found at the high-mass end of the FFP mass function ($\gtrsim$ a few $M_J$), where planet-planet scattering---which tends to liberate the lowest-mass planets present in systems---may efficiently remove planets only in the case of multi-super-Jupiter systems.

\subsection{Instabilities of circumbinary planets}
\label{subsection:p_type}

Well before the first circumbinary planetary system was discovered, numerical studies by \citet{dvorak1984numerical} \citet{dvorak1986critical}, and \citet{dvorak1989stability} established that there exists a critical circumbinary separation---the innermost semimajor axis for stability---beyond which planetary orbits may remain stable. Using a large suite of numerical simulations, the location of this innermost semimajor axis $a_c$ was empirically quantified by \citet{holman1999long} as 

\begin{equation}
\begin{split}
a_c/a_b = &(1.60 \pm 0.04) + (5.10\pm0.05)e \\ 
&+ (-2.22\pm0.11)e^2 + (4.12\pm0.09)\mu_b \\ 
&+ (-4.27\pm0.17)e\mu_b + (-5.09\pm0.11)\mu_b^2\\
&+(4.61\pm0.36)e^2\mu_b^2
\end{split}
\label{eq:ptype_stability}
\end{equation}
for systems of coplanar, circular orbits for binary semimajor axis $a_b$, binary mass ratio $\mu_b$, and binary eccentricity $e$. This stability boundary is visualized in red in Figure \ref{fig:stability_binary} for a binary with mass ratio $\mu_b=0.3$.

Later work that considered mutual inclinations between orbits demonstrated that islands of stability exist at high mutual inclinations for binaries on eccentric orbits. \citet{pilat2003stability} applied numerical experiments to produce stability maps as a function of binary separation for planets that orbit equal-mass binaries at mutual orbital inclinations of up to $i=50\degree$. This was followed by an analytic characterization of stability across binary orbital eccentricities and planetary inclinations, conducted by \citet{farago2010high} in the secular and quadrupolar approximations. Additional studies have demonstrated an island of stable libration toward polar orbits in circumbinary systems \citep{doolin2011dynamics, quarles2016long, li2016uncovering}. While polar circumbinary planets have yet to be definitively detected, \citet{martin2023ac} found that AC Herculis, which is a post-asymptotic giant branch star hosting a circumbinary disk, is consistent with this configuration.

More recent advances have extended this previous work by examining substantially larger-scale suites of numerical simulations \citep{quarles2018stability,georgakarakos2024empirical} and additional methods, such as deep neural networks \citep{lam2018machine}. These approaches were leveraged to more thoroughly delineate the parameter space for stability in circumbinary planetary systems, including islands of stability that are not well-captured by the polynomial form of Equation \ref{eq:ptype_stability}.

\subsection{Planetary orbital instabilities in multi-star systems}
While most non-single-star stability studies to date have focused on binaries, planets may similarly undergo orbital instabilities in triple-star or higher-order systems. There are many possible configurations of planetary orbits in such systems, including both $s-$type and $p-$type orientations.

A few systematic studies have considered such systems. \citet{verrier2007planetary} characterized the stability zones for planets within low-inclination hierarchical triple-star systems, demonstrating that the empirical results from \citet{holman1999long} hold for triple-star systems with relatively low orbital eccentricities and sufficiently wide-orbiting, low-mass stars. \citet{busetti2018stability} later applied $N-$body simulations to more extensively map planetary stability in triple-star systems, considering both prograde and retrograde planetary configurations. 

\subsection{Predictions and implications: instabilities in systems with bound stellar companions}
\label{subsection:binary_predictions}

Orbital instabilities in binary and multi-star systems do not always lead to planetary ejection. Instead, they may result in planetary or stellar collisions, or in some cases capture into different orbits: for example, a planet in a circumbinary orbit may be captured into a circumstellar orbit. 

Despite the extensive history of numerical simulation studies examining these systems, the distribution of these outcomes has not, to the authors' knowledge, been explicitly reported for binary systems with circumstellar planets. Instead, the systems are typically grouped only as ``stable'' vs. ``unstable'', without explicitly parsing the outcomes of the unstable systems. In circumbinary systems, by contrast, these outcomes have been explicitly traced and reported for some simulation suites.

\citet{smullen_planet_2016} considered a range of planet populations and binary configurations to demonstrate that, for various initial condition parameterizations in the literature, instabilities in circumbinary star systems tend to lead to ejection within 10 Myr, with ejections occurring at a far higher rate (typically $\approx50-70\%$) than in single-star systems (typically $\approx25-50\%$). \citet{sutherland_fate_2016} similarly conducted a numerical study of planets on tight (1-4 au) circular orbits in circumbinary systems with unequal stellar masses ($M_2/M_1=0.1$) and binary semimajor axis 1 AU, finding that roughly 80\% of unstable systems culminated in planetary ejection---with higher-eccentricity stellar binaries taking longer on average to eject planets. A small but nonzero percent of unstable systems in this simulation suite ($0.16\pm0.06\%$) had the unstable planet captured into orbit around the secondary star.

Intuition for the outcomes of instability in circumbinary systems may be gained through examining the circular restricted three-body problem \citep[see e.g.][]{szebehely1981stability}. For a three-body system setup with $m_1>m_2$ and $m_1, m_2 \gg m_3$, as well as circular orbits, the Jacobi constant $C_J$ is  conserved, given as

\begin{equation}
C_J = 2n(x\dot{y} - y\dot{x}) + 2\Big(\frac{\mu_1}{r_1} + \frac{\mu_2}{r_2}\Big) - (\dot{x}^2 + \dot{y}^2 + \dot{z}^2)
\end{equation}
for positions ($x, y,z$), velocities ($\dot{x}, \dot{y},\dot{z}$), binary mean motion $n$, and parameterized masses $\mu_1=Gm_1$ and $\mu_2=Gm_2$. In this regime, zero-velocity curves may be calculated for a given $C_J$. \citet{smullen_planet_2016} demonstrated that scattering is the only possible outcome for Jacobi constant $C_J>3.46$; that is, collisions may not occur for $C_J>3.46$. A wide swath of the instability region of circumbinary planets satisfies this condition, such that most systems that undergo instabilities in circumbinary systems are expected to undergo ejection, rather than collisions.

Using a combined disk evolution and $N-$body model designed to simulate the outcomes of core-accretion planet formation, \citet{coleman_properties_2024} found that free-floating planets ejected from circumbinary interactions have relatively high excess velocities 8-12 km/s, with velocity dispersions that are substantially higher ($\approx3\times$ larger) than those of free-floating planets ejected from planet-planet interactions. \citet{teasdale2026formation} simulated the process of disk fragmentation around circumbinary systems and found that orbital instabilities among the resulting population of planets---formed beyond 50 AU, and with a typical separation of 100 AU---would produce free-floating planets with lower ejection velocities, potentially making them more difficult to distinguish from planets ejected from scattering in single-star systems. Some giant ($\approx10M_J$) planetary-mass objects have been found on such wide circumbinary orbits \citep{christiansen2025nasa}, though their prevalence at a demographic level is currently not well-constrained.

Follow-up population synthesis work by \citet{coleman_predicting_2025} showed that instabilities in circumbinary systems may account for most of the inferred census of ejected high-mass ($>1M_{\oplus}$) FFPs. \citet{coleman_predicting_2025} offered quantitative predictions for the number density of ejected planets derived from orbital instabilities in single and binary star systems as a function of planet mass. Their results suggest that planetary ejection across the galactic population of stars should produce a trough in the mass distribution at roughly $0.8M_{\oplus}$---between the relatively high rate of very low-mass planets and the pebble isolation mass peak near $8M_{\oplus}$, above which fewer planets form and the number density of ejected planets turns back over.

Last, the relevance of an unstable region for planetary ejection implicitly relies on the formation or migration of planets into that region during the system lifetime. Do planets form or arrive in unstable regions in the first place? If planet formation is ubiquitous across wide swaths of parameter space in planetary systems, including these unstable regions, then large populations of FFPs may be produced through such instabilities. Otherwise, the census of ejected planets from such binaries may be much lower.

\section{Flyby-induced ejections}
\label{section:flybys}

\begin{figure*}
    \centering
    \includegraphics[width=1.0\textwidth]{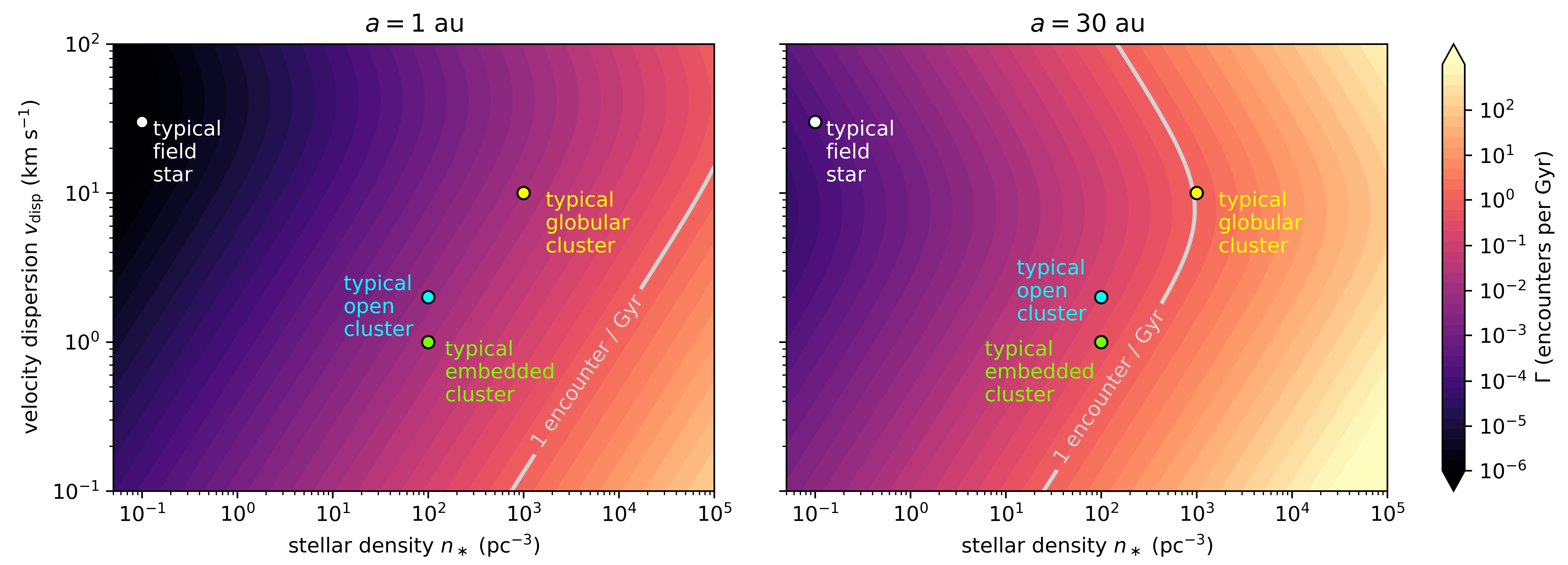}
    \caption{Flyby rates $\Gamma$ as a function of stellar density and velocity dispersion. Representative values for example stellar environments are shown alongside a reference line tracing the threshold for one flyby encounter per Gyr. We visualize two cases, for a planetary orbit with $a=1$ AU and another with $a=30$ AU. Wide-orbiting planets are much more susceptible to disruption by flybys than shorter-period planets due to their larger cross-sections for interaction.}
    \label{fig:flyby_rates}
\end{figure*}

Beyond intra-system dynamics, planets may also be dislodged from their primordial systems through dynamical interactions with passing neighbors that are not bound to the host star. A planet may be ejected from its host system in the case that the energy injected from the flyby perturber (or from the cumulative gravitational assists from several perturbers) overcomes the gravitational potential energy binding the planet to its host star. 

An extensive set of literature exists both analytically and numerically constraining the outcomes of flyby interactions for a planetary system \citep[e.g.][]{heggie1975binary, hills1984close, hills1989close,heggie1996effect,adams2001constraints,malmberg2011effects,li2015cross,brown2022on,kaib2025influence}. We do not attempt to provide a comprehensive overview of all possible outcomes of these flybys here. Instead, we focus on the occurrence of such encounters and relevant implications for planetary ejections. A useful overview of the relevant analytic formulae for the change in eccentricity and binding energy to two-body systems caused by flyby encounters is provided in Appendix A of \citet{spurzem_dynamics_2009}, which also draws upon prior results from \citet{heggie1975binary, heggie1996effect, roy2003energy, heggie2006few}.

In this section, we provide an overview of the expected rates of planetary ejection through close encounters with other stars and planets in the galaxy. We consider specific predictions for environments at a range of stellar densities.

\subsection{Flyby occurrence rates}
In general, the occurrence rate of stellar flyby interactions $\Gamma$ may be estimated as

\begin{equation}
\Gamma = \langle n_*\rangle \sigma\langle v\rangle,
\label{eq:flyby_rate}
\end{equation}
where $\langle n_*\rangle$ is the local mean stellar density, $\sigma$ is the cross-section of interaction, and $\langle v\rangle$ is the mean relative velocity between stars. As shown by this expression, planetary systems in dense stellar environments (high $\langle n_*\rangle$) and hosting wide-orbiting planets (large $\sigma$) are most susceptible to disruption by stellar flybys and subsequent planetary ejection. We show a mapping of $\Gamma$ as a function of stellar density and velocity dispersion in Figure \ref{fig:flyby_rates} for two example planetary orbit cross-sections.

\subsection{Cross-sections for flyby interactions}
\label{subsection:cross_sections}
At first glance, it may appear that the cross-section of interaction for a planetary system on a circular orbit would correspond to $\pi a^2$ for a given semimajor axis $a$: the system's geometric cross-section. This is not, however, generally the case due to the influence of gravitational focusing. 

Gravitational focusing, which describes how the gravitational influence between two bodies affects the effective cross-section for interaction, gives
\begin{equation}
\sigma = \pi a^2 \Big(1 + \frac{v_{\rm esc,*}^2}{v_{i}^2}\Big)
\label{eq:sigma_gravfoc}
\end{equation}
as the cross-section for a star interacting with a planet with semimajor axis $a$. Here $v_i$ is the relative speed of encounter between the two systems, which for stellar flybys may be estimated as the velocity dispersion $v_{\rm disp}$ of the background stellar population, while $v_{\rm esc,*}$ is the escape velocity from the planet's orbit around its host star. We note that this formalism applies specifically for close approaches to individual planets' orbits, whereas flybys at more distant close approaches may still significantly perturb the system. This formalism is conceptually very similar to that of the Safronov number $\Theta$ introduced in Section \ref{subsubsection:safronov}, which was originally derived through the framework of gravitational focusing to characterize the efficiency of planetesimal accretion \citep{safronov1969relative}.

Equation \ref{eq:escape_velocity_star} shows that the escape velocity from the stellar system scales as $v_{\rm esc, *}\propto 1/\sqrt{r}$, which can be approximated as $v_{\rm esc, *}\propto 1/\sqrt{a}$ for near-circular orbits. Combining this escape velocity with Equation \ref{eq:sigma_gravfoc} reveals that the cross-section for flyby interactions scales with the planetary semimajor axis as $\sigma\propto a$ for $(v_{\rm esc,*}^2/v_{i}^2)\gg 1$. This criterion is typically satisfied in open clusters, which have $v_{\rm disp}\approx1$ km/s, and for relatively short-period planets in globular clusters, with a higher typical velocity dispersion $v_{\rm disp}\approx10$ km/s (for comparison, $v_{\rm esc, *}=42, 19,$ and 8 km/s at Earth's, Jupiter's, and Neptune's orbit). \citet{li2015cross} verified this result through numerical simulations and demonstrated that the cross-section for disruption of a given planet's orbit is nearly independent of that of neighboring planets in the system. 

Field stars in the local solar neighborhood have typical relative velocities $v_{\rm disp}\approx30-40$ km/s \citep{nordstrom2004geneva} such that the approximation $(v_{\rm esc, *}^2/v_i^2)\gg 1$ breaks down particularly for wide-orbiting planets. In environments with very high velocity dispersions, such as the Milky Way bulge ($v_{\rm disp}\approx120$ km/s) and core ($v_{\rm disp}\approx170$ km/s), gravitational focusing plays a less dominant role and the geometric cross-section $\pi a^2$ is a more appropriate choice.

\subsection{Potential outcomes from stellar flyby interactions}

Depending on the timing of a flyby in the planetary system's evolution, as well as the relative masses and geometries of the flyby and planetary system, stellar flybys may result in several possible destabilizing outcomes. The implications of stellar flybys for planetary ejection have been examined in a range of contexts, including the solar system \citep{laughlin2000frozen, zink2020great,brown2022on,raymond2024future,kaib2025influence} and extrasolar systems \citep[e.g.][]{laughlin_modification_1998, malmberg2011effects, li2015cross, yu2024free}. Frequent stellar flybys in dense environments have observable implications for lower-mass bodies, as well, producing a significant population of ejected planetesimals that become interstellar objects \citep{hands2019fate, pfalzner2021significant}.

In stellar clusters, binary star systems may become unbound through either individual nearby encounters or through a more gradual process of ``evaporation'', in which bound objects are driven above the cluster's escape velocity by distant, weak encounters. Transposing these mechanisms to the planetary regime, \citet{delafuentemarcos1999runaway} showed that the evaporative process of ejection, while prevalent in binary star systems, is negligible for planets in single-star systems. Instead, using a series of $N$-body simulations, they found that planetary ejection in star clusters may occur through multi-star interactions that temporarily produce hierarchical triple or quadruple systems. This has the potential to launch newly unbound planets to velocities typically $<25$ km/s, with a small tail toward much higher values (up to $120$ km/s in the examined simulations). 

Close stellar flybys ($\lesssim100$ au) may induce the immediate ejection of neighboring planets or destabilize a system, leading to rapid ejection soon after the flyby event \citep{hills1984close}. In some cases, close flybys with low-mass stars or brown dwarfs may lead to an exchange interaction in which the planet becomes bound to the intruder \citep{malmberg2011effects}. 

More distant stellar flybys tend to excite the orbital eccentricities and inclinations of planets that are left bound, with the most pronounced effect on the widest-orbiting planets in a system \citep{malmberg2011effects, cai2018signatures, li2019fly, ellithorpe2022dynamical, rickman2023breakdown}. Eccentricity and inclination disturbances for wide-orbiting planets propagate inward through the system through secular planet-planet interactions \citep{zakamska2004excitation}, altering the orbits of closer-in planets. This reduces the timescale of orbital instabilities and planet-planet scattering, indirectly triggering planetary ejection over Myr to Gyr timescales \citep[e.g.][]{boley2012interactions}. Orbital excitation that occurs while the protoplanetary disk is still present may, however, be efficiently erased due to planet-disk interactions \citep{marzari2013circumstellar, picogna2014effects}. As a result, the long-term impact of dynamical excitation from stellar flybys is most pronounced post-disk-dispersal.

In rare cases with favorable orientations, a passing star with a close, edge-on encounter may induce the simultaneous ejection of two giant planets, producing Jupiter-mass binary objects \citep{wang2024free}. While such encounters are likely not common in the local solar neighborhood due to the low rate of close stellar encounters, the occurrence of such events may reach up to a few percent for initially wide-orbiting planetary systems residing in dense stellar clusters. Empirical upper limits on the occurrence of this mechanism may be placed by imaging surveys sensitive to binary planetary-mass objects \citep[e.g.][]{bouy2026multiplicity}. Importantly, however, many such objects may have been formed \textit{in situ}---through direct collapse---rather than through flyby-induced ejection (see Section \ref{subsection:relative_rates} for further discussion).

\subsection{Substellar object flybys}
Close flyby encounters with substellar objects (that is, free-floating brown dwarfs and planets) may also lead to the disruption of planetary orbits, as well as subsequent planetary ejection. Direct disruptions occur at a rate that can be calculated using an equivalent formalism to that presented in Equation \ref{eq:flyby_rate}, but with the local mean stellar density and relative velocity replaced with the relevant values for free-floating substellar objects, rather than stars. In the presence of an underlying population of FFPs, close encounters between FFPs and bound planetary systems may dislodge additional planets or induce orbital instabilities, further growing the FFP population. 

\citet{brown2025substellar} examined this scenario in the context of the solar system's mutual inclination and eccentricity excitation, finding from a suite of numerical simulations that close substellar object flybys often result in planetary ejection and could contribute $10^7-10^8$ free-floating planets ejected from the roughly $10^{10}$ sun-like star systems in the galaxy, assuming solar-system-like architectures. \citet{raymond2026was} considered a similar range of parameters to demonstrate that the flyby of a substellar object could have triggered a dynamical instability among the giant planets in the early solar system, finding a $\approx1-5\%$ chance of this scenario depending on the adopted occurrence rate of free-floating planets. 

The production of FFPs serves as a feedback loop that may enable the further production of FFPs. However, current projected rates of planetary ejection caused by substellar object flybys are relatively low: with only 1\% of sun-like stars having planets directly ejected through substellar object flybys, direct ejection through this mechanism likely does not play a major role in the total rate of FFP production. An enhanced instability rate from substellar object flybys offers an indirect avenue to increase the rate of planetary ejection through substellar object flybys, which may ultimately induce planet-planet instabilities. This rate, too, is limited by the occurrence of free-floating planets required for such encounters.

\subsection{Predictions and implications: flyby encounters}
The projected contribution of flyby encounters in producing FFPs varies significantly with stellar environment. The stellar density and velocity dispersion, each of which varies dramatically across the birth cycle and environment of a star, play an important role in setting the occurrence of stellar flybys, as can be seen through examination of Equation \ref{eq:flyby_rate}. Therefore, we consider predictions for planetary ejection specifically in stellar cluster environments due to their relatively high stellar densities, focusing on embedded clusters (Section \ref{subsubsection:embedded_clusters}), evolved open clusters (Section \ref{subsubsection:open_clusters}), and globular cluster environments (Section \ref{subsubsection:globular_clusters}).

\subsubsection{Predictions and implications: flybys in embedded clusters}
\label{subsubsection:embedded_clusters}
Stars are generally born in embedded clusters within molecular clouds, which are relatively high-density environments \citep[typically $10^2 - 10^3$ stars pc$^{-3}$;][]{lada2003embedded}. As a result, young stars experience a relatively high rate of stellar encounters that may trigger planetary ejection. 

In embedded cluster environments, protoplanetary disks may be truncated, warped, or destroyed by flyby encounters \citep{clarke1993accretion,scally2001destruction,adams2006early,fragner2009giant, forgan2009stellar,craig2013close,rosotti2014protoplanetary,zheng2015dynamical,cuello2020flybys,cuello2023close} and photoevaporation caused by Far Ultraviolet (FUV) and Extreme Ultraviolet (EUV) flux from nearby, high-mass stars \citep{storzer1999photodissociation, armitage2000suppression, scally2001destruction, adams2004photoevaporation, winter2018protoplanetary, parker2021external, winter2022external}, potentially suppressing the planet formation process. \citet{adams2006early} and \citet{proszkow2009dynamical} each found that these external effects should not prevent the formation of solar-system-like architectures in typical embedded clusters, though Jovian-mass planet formation may be suppressed in environments with the highest stellar densities \citep[$\rho\gtrsim10^4 M_{\odot}$ pc$^{-3}$;][]{daffern2022evaporation}. 

For environments in which giant planets are born, those residing in multiplanet systems and on the widest orbits are most susceptible to ejection. \citet{laughlin_modification_1998} found that Jupiter analogs commonly have their orbits disrupted by flyby interactions with binary star companions in their natal clusters, with roughly 5\% of such planets ejected over the course of 100 Myr. For solar-system-like architectures, with four gas giants, planetary ejection in embedded clusters occurs at a higher rate over the same timescale \citep[5-15\%;][]{malmberg2007instability} due to the combined influence of direct ejection and orbital-instability-inducing eccentricity excitation within multiplanet systems \citep[see also][]{hao2013dynamical}. Wider-orbiting giant planets at 10-100 AU may be ejected at much higher rates of up to tens of percent in embedded cluster environments \citep{parker2012,hao2013dynamical,fujii_survival_2019}.

Lower-mass planets may also be ejected through flybys in embedded clusters. \citet{chatterjee2012planets} considered the stability of \textit{Kepler}-detectable systems in open clusters, demonstrating that the short-period population of exoplanets accessible to the transit method should not generally be directly disrupted via cluster dynamics---though eccentricity excitation combined with planet-planet scattering may play an important indirect role in ejection rates. By comparing simulations of multi-Jupiter and multi-Earth systems at separations ranging from roughly 1-200 AU, \citet{cai2017stability} found that lower-mass wide-orbiting planetary systems have consistently higher survival rates, likely associated with their longer instability timescales (see e.g. Section \ref{subsubsection:amd}). In the case that every star forms with several planets spanning a wide range of masses and separations set by the \citet{kokubo1998oligarchic} oligarchic growth model, \citet{van2019survivability} found that flyby encounters in embedded clusters may produce 0.24-0.70 free-floating bodies per main-sequence star, spanning a wide mass range (0.01-130$M_J$; mean mass $1.4M_J$) that includes and extends beyond the planetary regime.

\subsubsection{Predictions and implications: flybys in evolved open clusters}
\label{subsubsection:open_clusters}
After leaving their natal molecular clouds, many stars remain in loosely bound, lower-density ($\leq10^2$ stars pc$^{-3}$) open clusters that slowly disperse through dynamical interactions with passing stars and the galactic tide. \citet{bonnell_planetary_2001} performed simulations of these older open cluster environments with a fiducial velocity dispersion of 2 km/s, which they used to show that planets with orbital separations $<10$ AU generally survive disruption by stellar encounters in open clusters over their lifetimes ($\approx$1 Gyr). As such, only the widest-orbiting planets are likely ejected directly from stellar encounters in evolved open clusters. Giant exoplanets in evolved open clusters have been discovered with the radial velocity \citep{brucalassi2014three, malavolta2016gaps} and transit methods \citep{meibom2013same, mann2016zodiacal}, all well interior to this orbital separation (at least in part due to limitations of these detection methods and available observational baselines).

Using Runge-Kutta orbital integrations in the restricted 3-body limit, \citet{smith_free-floating_2001} found a relatively high typical velocity of of planets liberated from equal-mass binary star systems in open clusters, peaking at around 6 km/s. Later $N$-body simulations by \citet{hurley2002free}, examining the ejection of Jupiter-sized planets, found a lower typical velocity of roughly 2 km/s, potentially due to their use of unequal-mass binaries and system masses following the more observationally-driven initial mass function (IMF) of \citet{kroupa1993distribution}.

\subsubsection{Predictions and implications: flybys in globular clusters}
\label{subsubsection:globular_clusters}
Globular clusters are high-density stellar environments with mean densities of $10^3$ stars pc$^{-3}$ and densities of up to $10^4-10^6$ stars pc$^{-3}$ in their cores. While their specific formation mechanisms remain under debate, the Milky Way's globular clusters are classically thought to have originated through major mergers of disk galaxies, multiphase in situ collapse with highly efficient star formation early in the galaxy's history, and/or the accretion of dwarf galaxies \citep{brodie2006extragalactic}. The extreme birth conditions of globular clusters enabled them to survive to the present-day despite their formation 10-13 Gyr in the past, whereas open clusters typically disperse well within 1 Gyr \citep{lada2003embedded}. 

Given the elevated stellar encounter rates of such high-density environments, \citet{sigurdsson1992planets} found that planets in globular clusters would typically be rendered unstable over the ages of their host systems, though follow-up work showed that planets on tight orbits ($a\lesssim0.3$ au) or those in less dense parts of globular clusters may more easily survive \citep{davies2001planets}. \citet{bonnell_planetary_2001} modeled planetary ejection in globular cluster environments, adopting a fiducial velocity dispersion of 10 km/s and showing that planets with $a\gtrsim1$ AU for stellar densities $>10^3$ stars pc$^{-3}$, or $a\gtrsim0.1$ AU for denser environments with $>10^4$ stars pc$^{-3}$, are typically ejected within the age of the globular cluster. \citet{hamers2017hot} more recently showed that orbital eccentricity excitation from frequent flybys in globular clusters commonly induces planetary ejection in some systems while triggering high-eccentricity migration toward tightly bound, close-in orbits in others, providing a candidate hot-Jupiter formation pathway.

An important underlying requirement for FFP production through stellar encounters is that planets must first form within the systems of interest. Very few exoplanets have been detected in globular clusters despite targeted searches \citep{gilliland2000lack, nascimbeni2012hst}, though the true implications for the occurrence of such planets remain under debate \citep[e.g.][]{nascimbeni2012hst, masuda2017reassessment}. Planets and planet candidates that have been found include the PSR B1620-26 b circumbinary planet \citep{thorsett1993psr, arzoumanian1996orbital, sigurdsson2003young} as well as the M62H planetary-mass companion (though its high density, $\rho=11$ g cm$^{-3}$, at a mass of a few Jupiter masses places its planetary nature in question;  \citet{vleeschower2024discoveries}), both orbiting pulsars.

An absence of detected planets could be interpreted as a natural outcome of rampant instabilities, or it could instead be indicative of formation limitations. In the case of globular clusters, the latter interpretation is supported by evidence from the demographic properties of single-star systems residing outside of globular cluster environments. Globular clusters are comprised of very low-metallicity stars, and demographic surveys for planets around similarly metal-poor stars in other environments have thus far discovered few to no planets \citep{fischer2005planet, boley2024first}. This suggests that the conditions for planet formation may not be as easily met in globular cluster environments. 

As introduced in Section \ref{subsubsection:embedded_clusters}, in cluster environments protoplanetary disks may be substantially truncated by photoevaporation from neighboring high-mass stars and tidal encounters from close flybys, which could suppress planet formation from its early stages. \citet{bonnell_planetary_2001} found that stellar densities must be relatively high, at $\gtrsim10^4-10^5$ stars pc$^{-3}$, for disruptive disk encounters to occur within typical timescales for planet formation. These conditions are possible for young globular clusters, such that planet formation may be suppressed in these environments.

In the case that planets do form regularly in globular clusters, how would the resulting FFPs be distinguished? \citet{smith_free-floating_2001} mapped the velocity distributions of FFPs produced through ejection in globular clusters, demonstrating that they would typically have higher velocities ($11-12$ km/s) by comparison with FFPs ejected from open clusters. \citet{kremer_probing_2019} carried out $N-$body simulations to show that up to $2\times10^7$ FFPs may be produced in the Galactic halo due to dynamical encounters and tidal loss in globular clusters, assuming that 10-50\% of stars in globular clusters begin with planets. This population of FFPs, while small relative to the full population, may be distinguished through its differing galactic distribution: most other mechanisms would produce FFPs primarily originating in the galactic plane, whereas globular-cluster-produced FFPs should instead follow the spherically isotropic distribution of globular clusters that populate the galactic halo.

\subsubsection{How often do planets remain bound to the host cluster after being liberated from their host star?}

After they become unbound from their host star, cluster-ejected planets may either remain within their natal cluster or join the underlying population of FFPs within the galaxy, unbound from any stellar system. \citet{hurley2002free} demonstrated that ejected planets may be liberated from dense cluster environments relatively slowly after being ejected from parent star systems toward the core of the cluster, drifting outward on timescales of 0.1-1 Gyr before joining the field population of FFPs. \citet{spurzem_dynamics_2009} similarly found that FFPs in clusters do not exit immediately; in their $N-$body cluster simulations, only $\approx10\%$ of FFPs typically escaped the cluster after a simulation time spanning a few million encounters between bound planets and passing stars. \citet{wang2015close} showed that planets liberated from their host cluster experience dozens of close encounters within 1000 AU prior to escape from the cluster's potential well, in some cases further perturbing bound systems or resulting in re-capture by another star.

\citet{parker2012} found that roughly 30\% of ejected Jovian-mass planets in their simulations become unbound from the birth cluster within 10 Myr. Planets that escape the cluster have relatively high velocities $5-10$ km/s, with lower velocities for planets ejected from wider orbits ($30$ au orbits vs. 5 au orbits in \citet{parker2012}). \citet{zheng2015dynamical} later applied $N-$body simulations to map out the range of outcomes from planetary disruption in cluster environments, showing the distribution of not only FFPs bound and unbound to the cluster, but also of star-planet systems that remain bound vs. ejected from the cluster. While some FFPs are ejected immediately and others remain bound longer in dispersing embedded clusters, all ejected FFPs in dispersing clusters ultimately become part of the background galactic population over $\approx$10 Myr timescales. 

Within stellar clusters, how are FFPs radially distributed? \citet{flammini2025dynamical} and \citet{flammini2026dynamical} applied $N-$body simulations to show that ejected low-mass bodies, such as planets and free-floating comets in clusters, should not undergo observable mass segregation in clusters in the same way that higher-mass bodies such as brown dwarfs and stars do. This implies that the population of FFPs that remain bound to a cluster would be potentially detectable throughout the cluster.

\section{Post-main-sequence planetary ejection}
\label{section:post-ms}

As a host star loses mass, the orbital properties of its surrounding planetary system correspondingly shift, in some cases making systems more susceptible to or directly inducing planetary ejection. In this section we consider how planetary orbits evolve in post-main-sequence systems, focusing on the implications for planetary ejection in two limits. 

The first of these is slow, adiabatic mass loss, as during the asymptotic giant branch (AGB) phase of stellar evolution. The second is effectively instantaneous mass loss, appropriate to describe planetary evolution during the explosion of a Type II (core-collapse) supernova when the host star exhausts its fuel and undergoes runaway gravitational collapse that causes the star to explode. We note that the impulsive regime examined in this work may also be applicable for other cases in which the orbital period far exceeds the mass-loss timescale, such as the evolution of some Oort cloud objects during the tip of the red giant branch. More detailed discussion of post-main-sequence orbital evolution may be found in the review by \citet{veras2016post}.

\subsection{The general variable-mass two-body problem}
Where mass loss is assumed to be isotropic, the equations of motion in the variable-mass two-body problem are quantified as \citep{omarov1962on,hadjidemetriou1963two, deprit1983secular,li2008influence,veras2011great, voyatzis2013multiplanet}

\begin{equation}
\mu \equiv M_* + M_p
\label{eq:mu_sum}
\end{equation}

\begin{equation}
\frac{da}{dt} = -\frac{a(1 + e^2 + 2e\cos f)}{1 - e^2} \frac{1}{\mu} \frac{d\mu}{dt}
\label{eq:dadt_postms}
\end{equation}

\begin{equation}
    \frac{de}{dt}=-(e+\cos f)\frac{1}{\mu} \frac{d\mu}{dt}
\end{equation}

\begin{equation}
    \frac{di}{dt}=\frac{d\Omega}{dt} = 0
\end{equation}

\begin{equation}
\frac{d\omega}{dt} = \frac{d\varpi}{dt} = -\frac{\sin f}{e} \frac{1}{\mu} \frac{d\mu}{dt}
\end{equation}

\begin{equation}
\frac{df}{dt} = -\frac{d\varpi}{dt} + \frac{n(1 + e\cos f)^2}{(1- e^2)^{3/2}}.
\end{equation}

These expressions describe the evolution of the orbit of a planet with mass $M_p$ orbiting a star with mass $M_*$, where $M_*$ is changing to produce a nonzero $d\mu/dt$. Here $i$ is the planetary orbital inclination, $\omega$ is the argument of periastron, $\varpi$ is the longitude of periastron, $\Omega$ is the longitude of ascending node, $f$ is the true anomaly, and $n$ is the mean motion $2\pi/P$ for orbital period $P$. The more general case of anisotropic and time-dependent mass loss was examined in \citet{veras2013exoplanet} (see Equations 34-38 in their work), which showed that the anisotropic terms' contributions scale as $\sqrt{a}$. 

\subsection{The mass-loss index $\Psi$}

For a constant mass-loss rate $\alpha$, the regimes of behavior may be described through a dimensionless ``mass-loss index'' $\Psi$ introduced by \citet{veras2011great} as
\begin{equation}
    \Psi \equiv \frac{\alpha}{n \mu}.
    \label{eq:mass_loss_index}
\end{equation}
Here $n=2\pi/P$ again corresponds to the planet's mean motion for orbital period $P$, while $\mu$ is the summed system mass (Equation \ref{eq:mu_sum}). The mass-loss index quantifies the relative timescales between the planet's orbital period and the rate of stellar mass loss. Where $\Psi\ll 1$ a system falls into the adiabatic regime, in which the planet's orbital period is much shorter than the mass-loss timescale. For effectively instantaneous mass loss in the impulse approximation, $\Psi\rightarrow\infty$.

\subsection{Direct ejection from slow mass loss}
We first consider the evolution of an individual planet's orbit in the slow, adiabatic mass-loss regime---where the orbital period is significantly shorter than the mass-loss timescale. In this scenario, the planet's orbital eccentricity remains roughly constant as the semimajor axis grows and the orbit gradually widens. The periastron $q$ increases monotonically with mass loss as \citep{veras2011great}

\begin{equation}
\frac{dq}{dt} = -\frac{a(1-e)(1-\cos f)}{1+e}\frac{1}{\mu}\frac{d\mu}{dt}
\label{eq:peri_change},
\end{equation}
which can be used to determine whether a planet's orbit is expanding more quickly than the radius of the stellar envelope. This can also be seen in Figure 2 of \citet{veras2011great}, which shows a sample orbital evolution trajectory in the adiabatic regime.

Eventually, the widening orbit's period becomes comparable to the mass-loss timescale ($\Psi\approx1$), transitioning the orbital evolution out of the adiabatic regime. When this transition takes place, $df/dt$ begins to librate rather than circulating as in the adiabatic regime, introducing rapid, runaway evolution of the orbital eccentricity and semimajor axis. The range of possible behavior is described in detail throughout \citet{veras2011great}, and we refer the reader to this work for a more in-depth overview. Here we describe only certain orbital limits to provide insight into the qualitative behavior of systems transitioning out of the adiabatic regime.

Where $f$ begins close to $0\degree$ or 180$\degree$ and $\Psi\gg 1$, libration in the true anomaly is small and thus $f$ can be treated as effectively constant. For $f=0^{\circ}$, this runaway is characterized as

\begin{equation}
e_{\rm runaway}|_{f=0^{\circ}} = e_0 \frac{\mu_0}{\mu} + \Big(\frac{\mu_0}{\mu} - 1\Big)
\end{equation}

\begin{equation}
a_{\rm runaway}|_{f=0^{\circ}} = \frac{a_0(1-e_0)}{2 - \frac{\mu_0}{\mu}(1 + e_0)}
\end{equation}
for initial semimajor axis $a_0$, initial eccentricity $e_0$, and initial summed mass $\mu_0$. In this regime, the eccentricity increases until the planet is ejected.

At the other extreme where $f=180^{\circ}$, the orbital eccentricity instead decreases toward $e=0$. Once the orbit has circularized, $|d\varpi/dt|$ grows and forces $df/dt$ to a nonzero value, changing $f$ such that the planet's orbit lands on another evolutionary track. $df/dt$ is nonzero for all true anomalies other than $f=0^{\circ}$, such that the value of $f$ will continue to change until the system reaches $f=0^{\circ}$, the eccentricity reaches the runaway growth track, and the planet is ultimately ejected. 

\subsection{Direct ejection in the impulse approximation}
We also consider the fast mass-loss limit in which all mass is lost instantaneously, corresponding to the impulse approximation. In this regime, $\Psi\rightarrow\infty$. The change in stellar mass from the initial value $\mu_i$ to the final value $\mu_f$ is characterized by the multiplicative factor $\beta$, given as

\begin{equation}
\mu_f = \beta \mu_i
\end{equation}
for $0<\beta\leq1$. In the impulse approximation, \citet{veras2011great} showed that a planet is ejected in the case that

\begin{equation}
\beta<\frac{1 + e_0^2 + 2e_0 \cos f_0}{2(1 + e_0\cos f_0)}
\label{eq:beta}
\end{equation}
where initial eccentricity is $e_0$ and initial true anomaly is $f_0$.

\begin{figure*}
    \centering
    \includegraphics[width=1.0\textwidth]{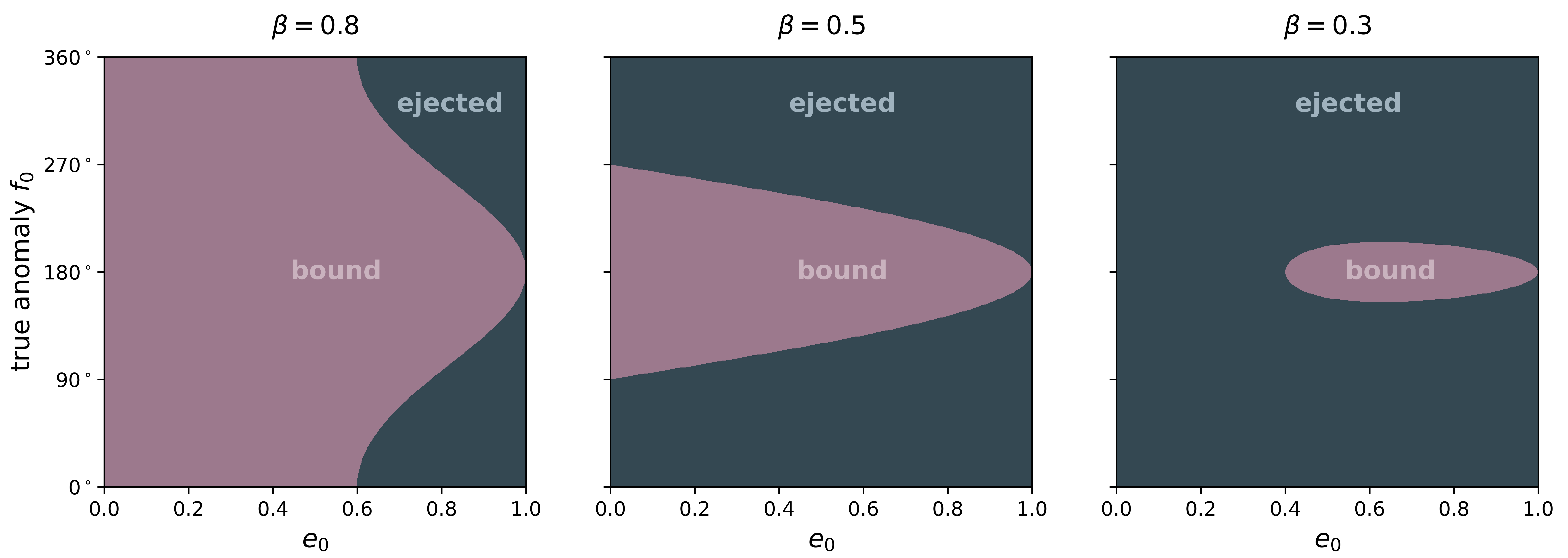}
    \caption{Planetary ejection regimes in the impulse approximation during a supernova explosion, shown for $\beta=0.8$, 0.5, and 0.3 where $\beta=\mu_f/\mu_i$ is the fractional stellar mass that remains after near-instantaneous mass loss. Here $e_0$ and $f_0$ are the initial orbital eccentricity and true anomaly of the planetary orbit. For higher fractional mass loss (increasing left to right across the three panels), planets are ejected across a wider range of true anomaly values. Note that for a planet to be ejected, it must survive the supernova without being destroyed by interactions with the stellar ejecta.}
    \label{fig:impulse_approximation}
\end{figure*}

We can see from Equation \ref{eq:beta} that, for $e_0=0$ and the limiting case $\cos f_0\rightarrow0$, more than half of the stellar mass must be lost instantaneously to eject a planet. For highly eccentric planets, depending on the true anomaly the planets may be more or less readily ejected from the system. Systems that most easily survive lie on highly eccentric orbits with $f_0\approx180^{\circ}$. Maps of the parameter space for bound and ejected planets are shown in Figure \ref{fig:impulse_approximation} for three example $\beta$ values.

\subsection{Post-main-sequence planetary close encounters}
\label{subsection:post-ms_close_encounters}
The orbital shifts induced by stellar evolution, in either the adiabatic regime or the impulse approximation regime, have the potential to trigger instabilities by pushing planetary orbits toward close approaches. As a result, the same planet-planet scattering instabilities described in Section \ref{section:close_approaches} may be hastened through the stellar evolution process, even in the case that planets are not directly ejected by the host star's evolution. 

Significant precedent from work leveraging $N-$body simulations has shown that planet-planet orbital instabilities are commonly triggered through the process of stellar evolution. Early work by \citet{duncan1998effects} used numerical simulations to show that the time to instability in the post-main-sequence solar system scales as a power law with the mass ratio between the sun and its neighboring planets, for ratios $\leq0.4$ times the current value. \citet{debes_are_2002} generalized this result to a broader range of planetary system architectures, demonstrating that two-planet systems that began marginally Hill stable \citep{Hill1886, gladman1993dynamics}---that is, stable against close approaches between planets---can become quickly susceptible to planet-planet instabilities through post-main-sequence stellar mass loss. 

The timescales of induced instabilities are often short relative to system lifetimes, such that they may have occurred in the past for many evolved star systems that hosted exoplanets while on the main sequence. \citet{voyatzis2013multiplanet}, for example, showed that instabilities in two-planet systems tend to occur in under 100 Myr for systems that undergo chaotic evolution due to stellar mass loss. \citet{mustill2014long} demonstrated that ejection-inducing instabilities are common for three-planet white dwarf systems, typically occurring at cooling ages of a few hundred Myr with ejections a factor of 3.9 times more likely during the white dwarf phase than during the host star's main-sequence phase. More recently, \citet{maldonado2022disentangling} carried out a suite of $N-$body simulations examining systems with 2-6 planets, demonstrating that the instability rate increases with planet multiplicity irrespective of the planet masses and range of separations, but that ejection in particular (rather than orbit-crossing events that do not result in ejection) is most efficient for systems of multiple high-mass planets. Observed short-period giant exoplanets found around red giant branch stars show orbital eccentricities effectively tracing the upper limit predicted for planet-planet scattering at a given orbital separation \citep{grunblatt2022tess}---a suggestive signature of planet-planet orbital instabilities in the past histories of evolved planetary systems.

\subsection{Increased susceptibility to flyby-induced ejection}

Another secondary planetary-ejection-related implication of stellar evolution is the increased flyby rate for surviving planets in an evolved system. Planetary orbits expand during post-main-sequence stellar evolution, such that the effective target area for a stellar or planetary flyby encounter also increases. This makes any given system that preserves its planets---that is, a system that does not have its full planetary system engulfed by the host star---more susceptible to planetary ejection through flybys. 

In the limit of slow, spherically-symmetric mass loss, planetary orbital separations evolve as $a\propto M_*^{-1}$ due to the approximate conservation of specific angular momentum. This means that a star losing half its mass will have its neighboring planets' orbits expand by a factor of 2 in the gravitational focusing regime ($\sigma\propto a$), or a factor of 4 in the geometric cross-section regime ($\sigma\propto a^2$; see Section \ref{subsection:cross_sections}). The close stellar encounter occurrence rate (Equation \ref{eq:flyby_rate}) increases by a commensurate factor.

\citet{zink2020great} showed that the sun's orbital expansion will enable close stellar encounters to ultimately (within 100-1000 Gyr) destabilize all of the outer solar system giant planets. This process, however, is slow: the first planet is typically ejected within $\approx30$ Gyr in simulations considered by \citet{zink2020great}. \citet{raymond2024future} found that individual flybys even within 100 AU would likely not destabilize any of the solar system planets: instead, multiple stellar flybys would generally be necessary to produce a sufficiently large change in acceleration to trigger planetary ejection. 

Flybys in post-main-sequence systems were examined in further detail and within a Galactic context by \citet{veras2014great}, who predicted a high rate of FFP production from stellar flybys in the galactic bulge. \citet{parker2026white} recently simulated the evolution of planets orbiting white dwarf progenitor stars with $1\leq M/M_{\odot}<2.5$ in dense ($10^4M_{\odot}$ pc$^{-3}$) embedded cluster environments, simultaneously modeling the influence of stellar evolution and $N-$body interactions. Incorporating the influence of orbit-widening from white dwarf evolution, \citet{parker2026white} showed that up to 48\% of their simulated planets orbiting white dwarfs, which include Jupiter-mass planets uniformly placed at 0.1-50 AU initial orbital separations, were disrupted to become FFPs. Their work predicts that many white dwarfs may subsequently capture these ejected FFPs, culminating in a population of wide-orbiting white-dwarf planets with semimajor axes $a>100$ AU and high orbital eccentricities $e>0.5$.


\subsection{White dwarf kicks}
Anisotropies---that is, deviations from perfect spherical symmetry---in mass loss during the AGB phase of stellar evolution produce a recoil kick of $\lesssim 2$ km/s on the stellar remnant, often referred to as a ``white dwarf kick'' \citep{fellhauer2003white,el2018imprints}. The dynamical influence of mass-loss anisotropies has been examined within the impulse approximation \citep{hills1983effects} as well as within the slower, adiabatic mass-loss regime \citep{hwang2025white,oconnor2026fate}. 

The implications of white dwarf kicks for planetary ejection have been most explicitly modeled in the impulse approximation, appropriate for wide binaries at separations $<1000$ au \citep{oconnor2026fate}. In this case, \citet{stephan2026contribution} leveraged $N-$body models together with the velocity distribution of white dwarf kicks derived in \citet{el2018imprints} to demonstrate that white dwarf kicks rarely trigger direct planetary ejection ($\lesssim1-2\%$ of known exoplanet systems). Instead, they tend to excite orbital eccentricities and inclinations that induce instabilities over time in multi-planet systems (see also Section \ref{subsection:post-ms_close_encounters}). \citet{stephan2026contribution} finds that such instabilities result in planetary ejection within 5 Myr for $\sim40-50\%$ of known exoplanet systems in which the planetary orbits would not be engulfed during the AGB phase of stellar evolution. Although \citet{stephan2026contribution} finds that FFPs produced by white dwarf kicks can reproduce only of order a few percent of the total FFP population, they also predict that such planets should slowly drift from their hosts at a rate of roughly 0.75 pc/Myr (0.73 km/s), offering a potential avenue toward identification.

\subsection{The role of binarity in planetary ejection for evolving stellar systems}
Up to this point, this section has focused primarily on planetary ejection in the case that an evolving single star is the planet host, unbound to any companion stars. The range of ejection scenarios is, however, more expansive when considering binary and higher-multiplicity systems, including influences from Type II supernovae of both stars, Type Ia supernovae, and common envelope evolution. 

These possibilities were mapped by \citet{veras_great_2012}, who delineated the critical semimajor axes beyond which planets became unbound as a function of binary parameters, metallicity, and changing mass loss rate. This work showed that circumbinary ($p-$type) planets are particularly susceptible to ejection during the post-main-sequence evolution of one or both stars in their system, with projected ejection rates that are highly dependent upon the binary properties and initial planet semimajor axes. \citet{nigioni2026quest} built upon this work by examining comparable two, three, four, and five-planet circumbinary systems. These authors showed that planetary ejection is common in such systems and that a modest sample of remnant, lower-multiplicity bound planets may be detectable with the future Laser Interferometer Space Antenna (LISA) gravitational wave observatory \citep{amaro2017lisa}.

While \citet{veras_great_2012} did not explicitly model evolving binary systems with $s-$type planets, they noted that planets in close binaries are likely to be destroyed by stellar evolution, whereas planets in wider binaries evolve more similarly to single-star systems with evolving stellar hosts. \citet{kratter_star_2012} showed that for planets on $s-$type orbits in binary star systems, stellar evolution may lead to an exchange of planets between stars for relatively large semimajor axis ratios $a_p/a_*\gtrsim0.1$, though their simulations were set up such that the minimum Jacobi constant reached by their simulations precluded ejection as a potential outcome. \citet{veras2017binary} later modeled the dynamical evolution of $s-$type two-planet systems in which the planet host undergoes stellar evolution with a bound stellar companion---showing that systems remain stable over 14 Gyr in the case that an $e=0$ binary companion is at least seven times as distant as the outer planetary orbit, with higher instability rates for more eccentric binaries.

\subsection{Predictions and implications: post-main-sequence planetary ejection}
While each of the aforementioned mechanisms has the potential to produce FFPs, direct planetary ejection from post-main-sequence stellar evolution likely plays a relatively minor role in producing FFPs. By evolving stars along their evolutionary tracks, \citet{veras2011great} showed that only planets on extremely wide orbits---typically at semimajor axes of hundreds to thousands of AU---are ejected during the AGB phase of stellar evolution. The same authors showed that nearly all planets that are not engulfed are ejected from orbit around a Type II supernova: the runaway regime is reached for all true anomalies except for a narrow range around $f_0=180^{\circ}$. However, these high-mass stars are uncommon, as the stellar IMF is heavily tilted toward lower-mass M stars.

Planets forming at the wide separations necessary for AGB ejection are likely rare, as protoplanetary disks typically have dust disk radii of roughly 30-45 AU \citep{tobin2020vandam}. Nevertheless, it is possible that a small fraction of planetary systems ($1-5\%$) retain very-wide-orbiting planets at orbital separations of a few hundred to a few thousand AU, trapped after near-ejection by dynamical instabilities during the embedded cluster phase. Some objects ranging from planet- to brown-dwarf-mass have been found at these wide orbital separations \citep[e.g.][]{luhman2011discovery,bailey2014hd,bohn2020two,janson2021wide,bohn2021discovery}.

Although only a small fraction of all stars undergo Type II supernovae, the population of FFPs produced by this mechanism may be distinguishable from other FFP sub-populations due to their high ejection velocities. \citet{regaly_lost_2022} modeled high-velocity (1,000-10,000 km/s) stellar envelope ejections and found a wide range of resulting planetary ejection velocities, ranging from $\approx1-275$ km/s and a mean ejection velocity of $\langle v_{\rm ejec}\rangle=18$ km/s.  

Secondary influences from post-main-sequence stellar evolution, including enhanced rates of orbital instabilities and disruptive stellar flybys, also contribute to FFP production. Given the low rate of FFPs expected directly from slow mass loss, together with the small fraction of stars that undergo Type II supernovae, these secondary effects likely represent the most dominant mechanisms for FFP production through post-main-sequence stellar evolution. 

\section{Discussion}
\label{section:discussion}

\subsection{Relative rates of FFP production}
\label{subsection:relative_rates}
As discussed in the previous sections, several planetary ejection mechanisms have been examined in detail, with differing FFP production rates across stellar environments. These mechanisms each contribute to the occurrence rate of free-floating planetary-mass objects. Therefore, FFPs as a whole will not follow any individual predicted distribution that focuses on only a subset of stellar populations. Instead, they will emerge with overlapping distributions of properties. 

We provide a high-level summary of planetary ejection mechanisms in Table \ref{tab:processes_summary} alongside the projected properties of FFPs produced by each mechanism. The efficiency of ejection is described under the assumption that all stars host planets following known limits on exoplanet demographics. We also report the typical projected excess ejection velocity $\langle v_{\infty} \rangle$ for each mechanism, noting that the velocity distributions are dependent on the system configuration at the time of ejection, with differing planet populations required for ejection across mechanisms. Several FFP production mechanisms have the potential to contribute a large fraction of the underlying population, including planet-planet scattering, embedded cluster encounters, and binary instabilities.

\begin{table*}[t]
\centering
\renewcommand{\arraystretch}{1.0}
\begin{tabular}{l l l l l l}
\toprule
{\bf Mechanism} & {\bf \% stars} & {\bf ej. rate} & {\bf typical planet ejected} & {\bf $\langle v_{\infty} \rangle$} & {\bf References}                \\
\midrule
\midrule

Planet-planet scattering & $100\%$ & up to 100\% & low-mass, wide-orbiting & 2-6 km/s & [1,2,3,4] \\ 
\midrule
$s-$type binary instability & 10\% & up to 100\% & any mass, wide-orbiting & ? & [3,4] \\
\midrule
$p-$type binary instability & 9\% & up to 100\% & any mass, close-orbiting & 2-12 km/s & [3,4,5,6] \\ 
\midrule
Embedded cluster encounters & 100\% & 30-70\% & any mass, wide-orbiting & 5-10 km/s & [7,14] \\
\midrule
Evolved open cluster encounters & 10\% & 10\% & any mass, wide-orbiting & 1-2 km/s & [8] \\
\midrule
Globular cluster encounters & 0.5\% & 80\% & any mass, wide-orbiting & 11-12 km/s & [9,10] \\
\midrule
Galactic tide + wide binaries & 2\% & 30-60\% & low-mass, wide-orbiting & 2-6 km/s & [11] \\
\midrule
AGB-phase stellar evolution & 5\% & $\lesssim$2\% & any mass, wide-orbiting & 0.5-1 km/s & [12,15,16] \\ 
\midrule 
Type II supernovae & 0.1-1\% & 100\% & any mass, any orbit & 18 km/s & [12,13] \\

\bottomrule
\end{tabular}
\caption{Summary of the relative prevalence of and predictions for planetary ejection mechanisms. ``\% stars'' refers to the percent of all stars that are susceptible to a given mechanism, and ``ej. rate'' refers to the percent of those systems that are predicted to eject at least one planet. The efficiency of planet ejection varies significantly with the locations and masses of formed planets; therefore, we describe the ``typical planet ejected'' in the fourth column. Typical excess velocities $\langle v_{\infty} \rangle$ are listed in the fifth column. We note that FFPs may populate different parts of parameter space: for example, those ejected from clusters do not all immediately join the latent background population, and planets ejected from encounters in globular clusters may populate the galactic halo. References are as follows: [1] \citet{bhaskar_properties_2025}, [2] \citet{guo_formation_2025}, [3] \citet{coleman_predicting_2025}, [4] \citet{offner2023origin}, [5] \citet{coleman_properties_2024}, [6] \citet{teasdale2026formation}, [7] \citet{parker2012}, [8] \citet{hurley2002free}, [9] \citet{smith_free-floating_2001}, [10] \citet{bonnell_planetary_2001}, [11] \citet{kaib2013planetary}, [12] \citet{veras2011great}, [13] \citet{regaly_lost_2022}, [14] \citet{hao2013dynamical}, [15] \citet{obrien2024wd}, [16] \citet{stephan2026contribution}}
\label{tab:processes_summary}
\end{table*}

To estimate rates in Table \ref{tab:processes_summary}, we adopt the assumption from \citet{coleman_predicting_2025} that planets on $p-$type orbits may form around binaries with separations of up to 3 AU. Considering binary instability rates from \citet{holman1999long,quarles2018stability}, we use planet-to-star semimajor axis ratio $a_c/a_b\lesssim0.1$ as a threshold for $s-$type binary instability and assume that planets are typically located at a maximum semimajor axis of 30 AU, such that circumstellar planets in binaries with $a_b\lesssim300$ AU may be prone to instabilities. We also assume that circumstellar planets cannot form unless the host stars are separated by at least a factor of three times the typical snow line separation $3$ AU, corresponding to binary separations of roughly $a_b\gtrsim10$ AU. Last, we consider only binaries with $a_b>1000$ AU as candidates for a substantial influence from the Galactic tide. 

\begin{figure*}
    \centering
    \includegraphics[width=1.0\textwidth]{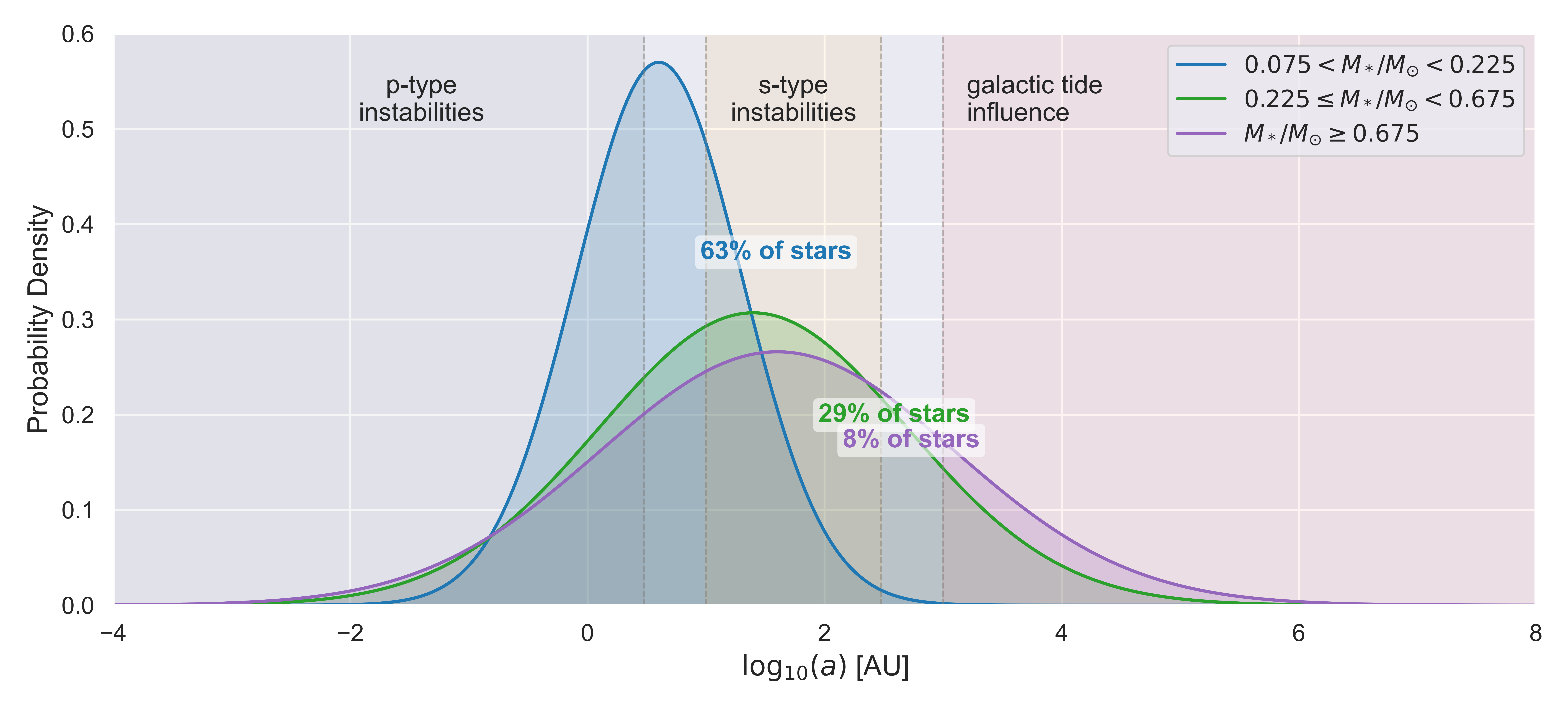}
    \caption{Log-normal binary separation distributions adopted from \citet{offner2023origin} and derived from data drawn from \citet{raghavan2010survey}. The adopted limiting regimes for three categories of binary-driven instabilities are shown as the shaded background regions. The percentage of all stars falling in a given mass range, derived from the \citet{kroupa2001imf} IMF, is shown together with the binary separation distribution for stars in that mass range that host a bound stellar companion.}
    \label{fig:separation_distributions}
\end{figure*}

Binary companion frequencies are drawn as a function of stellar mass from \citet{offner2023origin}, and we approximate that the same separation distributions extend over slightly wider mass ranges from $M_*=0.075-0.225M_{\odot}$, $M_*=0.225-0.675M_{\odot}$, and $M_*>0.675M_{\odot}$. The adopted separation distributions are shown in Figure \ref{fig:separation_distributions} for reference. Combining these fits with the piecewise power-law IMF from \citet{kroupa2001imf}, from which we find that 63\%, 29\%, and 8\% of stars fit into the listed low-, medium-, and high-mass bins, we find that 9\% of stars have a binary companion that may induce $p-$type binary instabilities if circumbinary planets form within the system, and 10\% of stars have a binary companion that may induce $s-$type binary instabilities for circumstellar planets. The true rate of initial planet formation in these unstable regions is not well-characterized, such that it is possible that all such systems eject one or more planetary companions. We find that 2\% of stars reside in binary systems that are sufficiently wide that they may be substantially influenced by the galactic tide ($a>1000$ au), potentially resulting in planetary ejection.

For open (globular) clusters, we consider densities of roughly $10^2$ ($10^3$) star pc$^{-3}$ and report ejection rates for planets that would be disrupted within 1 (10) Gyr at orbital separations 3 AU \citep{bonnell_planetary_2001}. Embedded cluster encounters may result in either a very low or very high rate of planetary ejection depending on the planetary multiplicity and primordial orbital separation distribution \citep{hao2013dynamical}; therefore, we report a range corresponding to outcomes for planets originating in multiplanet systems and spanning orbital separations of one to tens of AU.

We also assume that the galactic tide and stellar flybys in wide binaries typically induce planet-planet scattering comparable to that experienced by single stars, corresponding to a similar ejection velocity and source distribution of planets. The ejection rate reported in Table \ref{tab:processes_summary} for wide binary star systems corresponds to that of \citet{kaib2013planetary} with a four-giant-planet outer solar system analog and binary separations 1,000-30,000 AU. We note that the ``low-mass'' label within Table \ref{tab:processes_summary} refers to planets below the highest-mass planet within the system, but that planet-planet scattering---including that induced by galactic tide interactions---may still eject giant planets if multiple are present. 

Importantly, not all free-floating planets are the products of planetary ejection. Some FFPs may form \textit{in situ} as the lowest-mass tail of the stellar initial mass function \citep{luhman2012formation}, including from erosion of prestellar cores in the vicinity of high-energy radiation from OB stars \citep{whitworth2004formation,diamond2024formation}. Indeed, hundreds of planetary-mass objects have been discovered in star-forming regions since the turn of the century \citep[e.g.][]{oasa1999deep,lucas2000population,zapatero2000discovery,luhman2004new,scholz2009substellar,muzic2012substellar,luhman2016census,gagne2017banyan,robberto2020hst,gennaro2020hst,miret2022rich,pearson2023jupiter,luhman2024jwst,langeveld2024jwst,defurio2025identification}. Nevertheless, the high projected occurrence rate of low-mass FFPs \citep{sumi2023free}, together with new limits on the lowest-mass self-luminous objects in young star-forming regions---which taper off below roughly $3-5M_J$ \citep{langeveld2024jwst, defurio2025identification}---indicates that many FFPs, if they are truly unbound and free-floating, must have formed in and subsequently been ejected from circumstellar systems. 

Another important caveat is that apparent FFPs observed through microlensing are not necessarily unbound. Many of these bodies may instead be wide-orbiting bound planets, with stellar hosts at sufficiently large projected separations such that they were not captured by the microlensing event that identified the FFP. Thus, the true occurrence rate of FFPs retains significant ambiguity. Observations before or after a microlensing event, when the lens is tens to hundreds of mas in angular separation from its source, may be used to search for a potential host star and distinguish between these scenarios. This typically requires at least several years in temporal separation, considering lens proper motion rates at microlensing event distances. A small number of searches have been conducted to date using post-event follow-up imaging \citep{mroz2024free} and precovery searches with archival data \citep{kapusta2026hst}, with no significant detections of host stars to date.

Even with a high rate of dynamical instabilities, not all scattered bodies are fully ejected. \citet{gladman2002evidence} found that planet-planet scattering could push Mars-sized embryos to extremely wide orbits of a few hundred AU, and this scenario was revisited by \citet{silsbee2018producing} to further delineate the parameter space for planetary embryos that may remain bound to the solar system in the present day. \citet{scharf2009long} and \citet{veras2009formation} showed that even giant planets can be feasibly left on wide-orbiting, bound orbits at separations of hundreds of AU as the remnants of planet-planet scattering. In rare cases, scattered planets may be captured at orbital separations of thousands of AU as an Oort cloud population of exoplanets \citep{veras2009formation,bailey2019stellar, raymond2023oort, izidoro2025very}. \citet{hadden_free_2025} leveraged scattering simulations in conjunction with updated microlensing constraints to predict that about half of Neptune-mass bodies that appear as FFPs may be on wide, bound orbits, rather than being truly free-floating. While the mechanisms described in Table \ref{tab:processes_summary} \textit{can} account for all candidate FFPs, not all such objects should necessarily be accounted for as unbound products of planetary ejection. 

\subsection{Observed properties of candidate FFPs}
Discoveries of free-floating planets have been made for decades using direct imaging, and more recently using the gravitational microlensing method. Directly imaged FFPs in star-forming regions, all of which have been super-Jupiter-mass ($\geq3-5M_J$) to date \citep{langeveld2024jwst,defurio2025identification}, are generally consistent with formation through direct collapse with no need to invoke planetary ejection to produce the population. As a result, we focus in this section on the FFPs identified through gravitational microlensing. 

Ground-based gravitational microlensing surveys for free-floating and bound exoplanets include the Optical Gravitational Lensing Experiment \citep[OGLE;][]{udalski2015ogle}, the Korea Microlensing Telescope Network \citep[KMTNet;][]{kim2018kmtnet,henderson2014optimal}, and the Microlensing Observations in Astrophysics \citep[MOA;][]{hearnshaw2006moa} project. The threshold for FFP candidate events is typically placed at the bottom edge of the ``Einstein desert'', which corresponds to a dearth of microlensing events found with angular Einstein radii $10\,\mu\mathrm{as}\lesssim\theta_E\lesssim30\,\mu\mathrm{as}$ \citep{Kim2021, ryu2021kmt}. FFP candidates are generally reported for $\theta_E\lesssim10\,\mu$as, where the Einstein timescale $t_E$ and the angular Einstein radius $\theta_E$ are defined as 

\begin{equation}
t_E=\frac{\theta_E}{\mu_{\rm rel}}
\end{equation}
and

\begin{equation}
\theta_E\equiv\sqrt{\kappa M_L \pi_{\rm rel}}
\end{equation}
for 

\begin{equation}
\kappa\equiv\frac{4G}{c^2 \,\,\mathrm{au}}.
\end{equation}
Here $M_L$ is the lens mass, $\pi_{\rm rel}$ is the relative lens-source parallax, and $\mu_{\rm rel}$ is the lens-source relative proper motion. 

Roughly a dozen microlensing FFP candidates have been reported to date that are attributed to low-mass (super-Earth-mass or below) planets that are unlikely to have formed through direct collapse \citep{sumi2011unbound,mroz2017no,Mroz2018,Mroz2019,Mroz2020a,Mroz2020b,Kim2021,ryu2021kmt,Koshimoto2023,jung2024kmt,poleski2025kmt, Inyanya2026,dong2026free,ryu2026kmt}. The completeness-corrected high occurrence of these isolated objects has been leveraged to argue for a high rate of FFPs particularly at the low-mass end, with estimates from \citet{sumi2023free} corresponding to a total of $80^{+73}_{-47}M_{\oplus}$ ejected per star for objects in the mass range $0.33<M/M_{\oplus}<6660$.

The history of any individual FFP ranges from difficult to impossible to definitively determine. This is for the same reasons that it is difficult to definitively show the origin system of the interstellar object 'Oumuamua, as discussed by \citet{morbidelli2020no}: the processes that produce FFPs are chaotic, and it is impossible to integrate an object's trajectory backward in time to determine its origin (which would violate the second law of thermodynamics). In combination with the short dynamical lifetimes of free-floating bodies in the Galaxy, this means that orbital paths cannot be directly traced backward to origin systems unless the objects were ejected in the recent past. Some efforts have been made to identify possible progenitor systems for observed interstellar objects under the assumption of recent ejection, with plausible hosts reported for each discovered interstellar object \citep{zhang2018prospects,feng2018oumuamua,dybczynski2018investigating,bailer2018plausible,portegies2018origin,zuluaga2018general,bailer2020search,portegies2021oort,guo2025search}.

Some information may still be gleaned from the velocities of FFPs, even if their specific histories are difficult (or impossible) to unambiguously disentangle. In most cases, microlensing events associated with FFPs do not correspond to directly measured masses; instead, microlensing measurements provide $t_E$, which is a degenerate combination of  the true mass, the object velocity, and the distance to the lensing body. The measurement of finite-source effects \citep{witt1994can} can break one of these degeneracies; however, an intrinsic degeneracy typically remains. This final degeneracy was broken in one recent case reported by \citet{dong2026free}, in which combined observations from ground-based telescopes (OGLE and KMTNet) and the \textit{Gaia} spacecraft provided a parallax measurement of the event alongside measurements of $t_E$ and $\theta_E$. The most favored solution for the system corresponds to lens transverse velocities $v_{l,l,\rm VSR}=-48^{+23}_{-41}$ km/s and $v_{l,b,\rm VSR}=79\pm12$ km/s relative to the vicinity standard of rest (VSR) describing neighboring stars. While the velocity deviation of the FFP from the VSR is consistent with the Galactic velocity distributions within $2\sigma$, the potentially high observed velocity of this object is suggestive of ejection through a mechanism that can impart high relative velocities, such as instabilities in a tight circumbinary system or a Type II supernova explosion.

\subsection{Sub-planetary-mass ejected bodies}
In addition to planetary-mass ejected bodies, lower-mass free-floating objects are, in many cases, subject to the same ejection mechanisms described within this review. Understanding the properties of sub-planetary-mass ejected bodies may therefore offer useful insights into the ejection rates and mechanisms relevant for the FFP population. We do not attempt to provide a comprehensive overview of results oriented around the ejection of these sub-planetary-mass objects here. Instead, we provide a few key points to contextualize the lower-mass ejected population in light of current constraints on planetary-mass objects' ejection.

In the past decade, the first three interstellar objects---1I/'Oumuamua \citep{williams2017minor,meech2017}, 2I/Borisov \citep{Borisov2019CBET4666,jewitt2019initial,guzik2020initial}, and 3I/ATLAS \citep{denneau2025atlas,bolin2025interstellar,seligman2025discovery}---were discovered. These objects have radii of $\sim0.1-3$ km \citep{meech2017,jewitt2017interstellar,knight2017rotation,drahus2018tumbling,jewitt2020nucleus,jewitt2023interstellar,jewitt2025hubble}, and they were discovered passing through the solar system on hyperbolic orbits, unbound to the sun.

Several studies have explored the ejection of small bodies from their host planetary systems, largely with the goal of explaining the demographics of interstellar objects \citep[for a review, see e.g.][]{issi2019oumuamua}. These studies have made several key points.  First, the known exoplanet population, which has been largely discovered by the transit and radial velocity methods, contains relatively few planets capable of efficiently ejecting small bodies \citep{laughlin2017consequences,rice2019hidden}. Nevertheless, statistical evidence from planet searches that are less biased toward short-period planets, including suggestive evidence drawn from protoplanetary disk gaps in \citet{zhang2018disk} and microlensing searches in \citet{suzuki2016exoplanet}, indicates that wide-orbiting, Neptune-mass planets---efficient interstellar object ejectors \citep{rice2019hidden}---are common. Second, comet-like planetesimals are far more likely to be ejected than asteroid-like ones \citep{raymond2018implications}. This result stems from the expectation that rocky planetesimals are found much closer to their host stars, where they are typically accreted rather than ejected. Beyond the snow line there is also far more space to form planetesimals, for simple geometric reasons, enabling the production of a large population of icy interstellar object seeds. Third, planetesimals may undergo significant processing on their pathway to ejection, including tidal disruption and thermal processing \citep{cuk2018oumuamua,raymond2018interstellar,zhang2020tidal}.  

Like FFPs, interstellar objects may be produced by mechanisms other than scattering in single-star systems---including ejection from close binary star systems \citep{jackson2018ejection} and gravitational stripping of distant objects during stellar flybys \citep{vincke2016cluster,portegies2018origin,hands2019fate}. The projected velocity distributions of interstellar objects may in principle be differentiated between formation scenarios: \citet{pfalzner2021significant} modeled differences in interstellar object velocities for the cases of production by stellar evolution (typically $<0.5$ km/s), flybys (typically $<1$ km/s), and planet-planet scattering (typically above 2-3 km/s).  Ejection from circumbinary systems would likely produce interstellar objects at even higher velocities \citep{coleman_properties_2024}.

Unlike interstellar objects, the detection of free-floating exomoons remains elusive. Nevertheless, useful limits may be placed in the near future on moon-mass objects that remain bound to their host stars. Microlensing observations with the NASA \textit{Nancy Grace Roman} space telescope \citep{spergel2015wide, akeson2019wide} are projected to enable the first population-level studies of moon-mass, bound objects \citep{penny2019predictions} down to a Ganymede mass (0.025$M_{\oplus}$), demonstrating the rate at which sub-Earth-mass bodies persist on wide orbits without having undergone either collisions or ejections. Considering true solar-system analog exomoons, projections indicate that of order one bound exomoon with moon-planet mass ratio $10^{-4}-10^{-2}$ around a wide-separation ($0.3-30$ au) giant ($30M_{\oplus}-10M_{\rm Jup}$) planet will be detectable with the current \textit{Roman} observing strategy \citep{lastovska2025predictions}. \citet{johnson2020predictions} and \citet{derocco2026prospects} each showed that \textit{Roman} will be sensitive to free-floating objects down to roughly the mass of Mars (0.1$M_{\oplus}$), with sensitivity that drops rapidly at lower masses. Some FFPs detected by \textit{Roman} may have originated as moons if a subset of exoplanets hosts moons more massive than those in the solar system. For such low-mass free-floating bodies, it will likely be difficult to distinguish between objects that were initially moons and those that were planetesimals in their natal systems.

\subsection{Future prospects}
\label{subsection:future_prospects}

We stand at a key inflection point, with rich prospects for future observations from which we may glean insights into the processes underlying planetary ejection. Upcoming space missions with dedicated surveys amenable to microlensing FFP searches have the potential to provide unprecedented observational constraints on the properties of FFP populations. 

The NASA \textit{Nancy Grace Roman} space telescope, with projected launch in late 2026, offers tremendous prospects for upcoming microlensing detections of FFPs. Hundreds of FFPs are projected to be discovered by the mission's Core Community Surveys \citep{johnson2020predictions}, which would increase the number of known FFPs by a factor of ten. With a sufficiently large census of planets discovered to support population studies, the FFPs discovered by \textit{Roman} are projected to enable distinctions between current, competing models for the FFP mass function \citep{derocco2026prospects}. 

The Earth 2.0 mission, led by the Shanghai Astronomical Observatory through the Chinese Academy of Sciences for projected launch in late 2028, is also expected to find hundreds of FFPs during its four-year primary mission. Notably, this includes roughly 150 FFPs for which true masses will be measured from simultaneous parallax measurements through synergistic observations with KMTNet \citep{ge2022et}. These parallax measurements offer an unprecedented opportunity to characterize the transverse velocities of the lenses at scale, as in \citet{dong2026ffp}, for comparison with excess velocities imparted through distinct planetary ejection models.

On the theoretical side, inroads that would be particularly useful in advancing planetary ejection include clear formation-informed predictions for the planetary mass function at wide separations. How commonly do low-mass planets form at and beyond the ice line of a protoplanetary disk, where planetary ejection is efficient? How often do planets form on wide orbits in $s-$type binary star systems and in dense stellar cluster environments? Progress has been made in connecting microlensing, direct imaging, and ALMA disk constraints with the stellar and planetary mass functions \citep{yee2025microlensing}, and further such limits offer a pathway toward determining what fraction of FFP candidates are truly unbound.

The observed occurrence of FFPs provides an opportunity to characterize the total mass budget available for planet formation---which the classic minimum-mass solar nebula \citep{weidenschilling1977distribution} and its extrasolar analog \citep{chiang2013minimum} would generally underestimate relative to the population of free-floating planets. Heavier Class 0/I disks may be necessary to invoke to account for the high observed mass budget, if it is confirmed with upcoming observations \citep{lee2026persistent}. Constraints on the initial mass budgets of planetary systems have the potential to further inform the relative rates with which contributions from individual planetary ejection mechanisms may be capped.

\section{Summary}
\label{section:summary}

Planetary ejection is ubiquitous across environment, stellar multiplicity, and the life cycle of a stellar system: where planets form, planets are ejected. For a planetary system to form, its building blocks must have undergone instabilities, leading to orbit crossing and subsequent collisions through which planets grow. This process does not end with the formation of planets. In some cases, these continued instabilities may culminate in planetary ejection. 

In this review, we have discussed the dynamical mechanisms relevant for planetary ejection, including both underlying principles and the current state of the field. We provide some high-level summary points as follows.

\begin{itemize}
    \item Roughly a dozen candidate low-mass, sub-Jovian FFPs have been discovered by microlensing surveys. When incorporating survey completeness metrics, this discovery rate corresponds to a vast background population of low-mass FFPs. While higher-mass, directly-imaged FFPs---often found in star-forming regions---may form \textit{in situ} through direct collapse, lower-mass FFPs demand an alternative explanation. This implies a high rate of planetary ejection.
    \item Planets may be ejected from their host systems through dynamical mechanisms including planet-planet scattering, binary-induced instabilities (including resonance-overlap-driven orbital evolution, as well as eccentricity excitation by passing stars and the galactic tide), stellar and planetary flybys, and post-main-sequence orbital evolution. All of these mechanisms make some contribution to the population of FFPs.
    \item Though multiple mechanisms play a role in producing the FFP population, planet-planet scattering, embedded cluster encounters, and binary instabilities are likely predominant in FFP production. The true prevalence of planetary ejection via each mechanism examined in this work is set by the initial mass and semimajor axis distribution of planets forming across stellar environments.
    \item Radial-velocity and transiting exoplanet observations have revealed a prevalence of highly eccentric orbits among observed giant planets, together with architectural trends such as the dense packing of planetary systems on the brink of instability, that are suggestive of past planetary system instabilities. At a population level, these architectural vestiges may be the signposts of previous planetary ejection events.
    \item Upcoming microlensing surveys have the potential to place limits on the planetary initial mass function from the occurrence of FFPs while also improving constraints on the occurrence of wide-orbiting, sub-Jupiter-mass planets. In these ways, among others, these surveys will inform the characteristics of both FFPs and their potential parent populations.
\end{itemize}

Understanding the dynamics of planetary ejection requires an overarching synergy of techniques. It is informed by the occurrence of mature long-period planets that remain bound after planetary ejection, measured through radial velocity and astrometry surveys. It is informed by the transit-identified dearth of resonant chains, and by the fingerprints of those few pristine remnant resonant systems that have evaded instability. It is informed by the prevalence of young, wide-orbiting giant planets from interferometric protoplanetary disk imaging, by direct-imaging limits in the cradles of planet formation, and by the discovery of FFPs and the demographics of wide-orbiting planets offered through microlensing.

Each nexus in this web of constraints provides a small piece of the broader puzzle of planetary ejection. Planetary ejection extends to the heart of fundamental questions about how planetary systems form and evolve: how are planets distributed at birth? In which environments do they form? And in what ways does the current set of bound systems differ from the demographics at various points in the planetary systems' life cycles? 

These are large-scale problems, yet they are remarkably tractable. Significant, substantive progress has been made and remains to be made in coming decades---from the discovery of the first exoplanets only a few decades ago to the advent of ground- and space-based exoplanet-oriented surveys today. It has become possible to map not only the demographics of planetary systems, but also the demographics of the \textit{now-invisible}, past states of planetary systems. A complete picture of planetary systems' life cycles---at least the fuzzy outline, if not the fully sharpened image---may not be so far out of reach. 

\begin{acknowledgments}
We are grateful to the organizers of the 2025 Rogue Worlds Two conference, through which the core content of this article was originally presented as a review talk. We also thank the anonymous reviewer for helpful feedback that has strengthened this manuscript. M.R. acknowledges support from the Heising-Simons Foundation through Grants \#2021-2802 and \#2023-4478.  S.N.R. is grateful to the ANR's PEPR-Origins and CNRS/INSU's PNP programs for support. W.D. was supported by NSF grant PHY-2210361 and the Maryland Center for Fundamental Physics. 
\end{acknowledgments}

\begin{contribution}
M.R. was responsible for formulating and writing the manuscript. W.D. and S.N.R. each provided feedback on and contributed ideas toward refining the manuscript.



\end{contribution}

\software{\texttt{astropy} \citep{astropy2013,astropy2018,astropy2022}, \texttt{astroquery} \citep{ginsburg2019astroquery}, \texttt{matplotlib} \citep{hunter2007matplotlib}, \texttt{numpy} \citep{oliphant2006guide, walt2011numpy, harris2020array}, \texttt{pandas} \citep{mckinney2010data}, \texttt{scipy} \citep{virtanen2020scipy}}

%




\bibliography{bibliography_dec2025}{}
\bibliographystyle{aasjournalv7}



\end{document}